\documentclass[12pt]{article}
\usepackage{hyperref}
\usepackage{epsfig}
\usepackage{comment}
\usepackage{latexsym}
\usepackage{color}
\usepackage{amsmath}
\usepackage{amssymb,amsfonts}

\newcommand{\mysquare}[0]{\raise-.2ex\hbox{{\Large$\Box$}}}
\def\lsim{\mathrel{\rlap {\raise.5ex\hbox{$ < $}}
{\lower.5ex\hbox{$\sim$}}}}
\def\gsim{\mathrel{\rlap {\raise.5ex\hbox{$ > $}}
{\lower.5ex\hbox{$\sim$}}}} \topmargin -1.5cm \textheight=22.5cm \textwidth=16.5cm
\catcode`\@=11
\newcount\hour
\newcount\minute
\newtoks\amorpm
\hour=\time\divide\hour by60 \minute=\time{\multiply\hour by60 \global\advance\minute by-\hour}
\edef\standardtime{{\ifnum\hour<12 \global\amorpm={am}%
        \else\global\amorpm={pm}\advance\hour by-12 \fi
        \ifnum\hour=0 \hour=12 \fi
        \number\hour:\ifnum\minute<10 0\fi\number\minute\the\amorpm}}
\edef\militarytime{\number\hour:\ifnum\minute<10 0\fi\number\minute}
\def\draftlabel#1{{\@bsphack\if@filesw {\let\thepage\relax
   \xdef\@gtempa{\write\@auxout{\string
      \newlabel{#1}{{\@currentlabel}{\thepage}}}}}\@gtempa
   \if@nobreak \ifvmode\nobreak\fi\fi\fi\@esphack}
        \gdef\@eqnlabel{#1}}
\def\@eqnlabel{}
\def\@vacuum{}
\def\draftmarginnote#1{\marginpar{\raggedright\scriptsize\tt#1}}
\def\draft{\oddsidemargin -.2truein
        \def\@oddfoot{\sl preliminary draft \hfil
        \rm\thepage\hfil\sl\today\quad\militarytime}
        \let\@evenfoot\@oddfoot \overfullrule 3pt
        \let\label=\draftlabel
        \let\marginnote=\draftmarginnote
   \def\@eqnnum{(\theequation)\rlap{\k

 ern\marginparsep\tt\@eqnlabel}%
\global\let\@eqnlabel\@vacuum}  }
\newcommand{\be}[0]{\begin{equation}}
\newcommand{\ee}[0]{\end{equation}}
\newcommand{\bea}[0]{\begin{eqnarray}}
\newcommand{\eea}[0]{\end{eqnarray}}
\newcommand{\ba}[0]{\begin{eqnarray}}
\newcommand{\ea}[0]{\end{eqnarray}}
\def\bs{\begin{subequations}}
\def\es{\end{subequations}}

\def\thebibliography#1{%
\vskip 0.5cm \centerline{\bf \Large References}
\list{%
[\arabic{enumi}]}{\settowidth\labelwidth{[#1]} \leftmargin\labelwidth \advance\leftmargin\labelsep
\usecounter{enumi}}
\def\newblock{\hskip .11em plus .33em minus .07em}
\sloppy\clubpenalty4000\widowpenalty4000 \sfcode`\.=1000\relax}

\renewcommand{\theequation}{\arabic{section}.\arabic{equation}}

\renewcommand{\section}{\setcounter{equation}{0}\@startsection
{section}{1}{0mm}{-\baselineskip}{0.5\baselineskip} {\normalfont\Large\bfseries}}

\renewcommand{\subsection}{\@startsection
{subsection}{2}{0mm}{-\baselineskip}{0.5\baselineskip} {\normalfont\large\bfseries}}

\renewcommand{\subsubsection}{\@startsection
{subsubsection}{3}{0mm}{-\baselineskip}{0.5\baselineskip} {\normalfont\normalsize\slshape}}

\usepackage{amssymb,amsfonts}
\usepackage{graphicx}
\usepackage{cite}

\newcommand{\dis}{\displaystyle}

\newcommand{\Z}{\mathbb{Z}}
\newcommand{\Ka}{K{\"a}hler }
\renewcommand{\O}{{\cal O}}
\renewcommand{\Re}{{\rm Re}}
\renewcommand{\Im}{{\rm Im}}

\newcommand{\abs}{|}

\newcommand{\ie}{{\em i.e. }}
\newcommand{\where}{\mbox{where}}
\newcommand{\with}{\mbox{with}}

\renewcommand{\and}{\mbox{and}}

\newcommand{\N}{{\cal N}}
\newcommand{\M}{{\cal M}}

\newcommand{\F}{{\cal F}}

\renewcommand{\o}{\overset{\circ}}
\newcommand{\oo}{\overset{\circ\circ}}
\newcommand{\dil}{\phi_{d}}

\newcommand{\tk}{\tilde k}
\newcommand{\tm}{\tilde m}
\newcommand{\tn}{\tilde n}

\def\ab{[{}^a_b]}

\begin{document}
%\verb|\usepackage{draftcopy}|\\
\begin{titlepage}
\begin{flushright}
CPHT--L094.0909,
LPTENS--09/31,
October 2009
\end{flushright}

\vspace{1mm}

\begin{centering}
{\bf\huge Thermal and quantum induced early superstring cosmology}\\
%\vspace{3mm}
%{\bf\huge }\\

\vspace{6mm}
 {\Large F. Bourliot$^{1}$, J. Estes$^{1}$, C. Kounnas$^{2}$ and H. Partouche$^1$}

\vspace{2mm}

$^1$  {Centre de Physique Th\'eorique, Ecole Polytechnique,$^\dag$
\\
F--91128 Palaiseau cedex, France\\
{\em Francois.Bourliot@cpht.polytechnique.fr} \\
{\em John.Estes@cpht.polytechnique.fr}\\
{\em Herve.Partouche@cpht.polytechnique.fr}}

\vspace{2mm}

$^2$ Laboratoire de Physique Th\'eorique,
Ecole Normale Sup\'erieure,$^\ddag$ \\
24 rue Lhomond, F--75231 Paris cedex 05, France\\
{\em  Costas.Kounnas@lpt.ens.fr}

\end{centering}

%\begin{quote}
%\vspace{2mm}
$~$\\
\centerline{\bf\Large Abstract}
\vskip .1cm
\noindent
In this work, we review the results of Refs \cite{Cosmo-0} --\cite{Cosmo-4} dedicated to the description of the early Universe cosmology induced by quantum and thermal effects in superstring theories. The present evolution of the Universe is described very accurately by the standard $\Lambda$-CDM scenario, while very little is known about the early cosmological eras.  String theory provides a consistent microscopic theory to account for such missing epochs. In our framework, the Universe is a torus filled with a gas of superstrings. We first show how to describe the thermodynamical properties of this system, namely energy density and pressure, by introducing temperature and supersymmetry breaking effects at a fundamental level by appropriate boundary conditions.  

We focus on the intermediate period of the history: After the very early ``Hagedorn era" and before the late electroweak phase transition. We determine the back-reaction of the gas of strings on the initially static space-time, which then yields the induced cosmology. The consistency of our approach is guaranteed by checking the quasi-staticness of the evolution.
It turns out that for arbitrary initial boundary conditions at the exit of the Hagedorn era, the quasi-static evolutions  are universally attracted to radiation-dominated solutions. It is shown that at these attractor points, the temperature, the inverse scale factor of the Universe and the supersymmetry breaking scale evolve proportionally.  There are two important effects which result from the underlying string description.  First, initially small internal dimensions can be spontaneously decompactified during the attraction to a radiation dominated Universe. Second, the radii of internal dimensions can be stabilized.
%\end{quote}

\vspace{3pt} \vfill \hrule width 6.7cm \vskip.1mm{\small \small \small
  \noindent
$^\dag$\ Unit{\'e} mixte du CNRS et de l'Ecole Polytechnique,
UMR 7644.}\\
 $^\ddag$\ Unit{\'e} mixte  du CNRS et de l'Ecole Normale Sup{\'e}rieure associ\'ee \`a
l'Universit\'e Pierre et Marie Curie (Paris 6), UMR 8549.

\end{titlepage}
\newpage
\setcounter{footnote}{0}
\renewcommand{\thefootnote}{\arabic{footnote}}
 \setlength{\baselineskip}{.7cm} \setlength{\parskip}{.2cm}

\setcounter{section}{0}

%%%%%%%%%%%%%%%%%%%%%%%%%%%%%%%%%%%%%%%%%%%%%%
%%%%%%%%%%%%%%%%%%%%%%%%%%%%%%%%%%%%%%%%%%%%%%

% \tableofcontents
\section{Why and how studying superstring cosmology?}

We are aware of the existence of four interactions: Gravity, weak interaction, electromagnetism and strong interaction. Though gravity is the oldest known one, it is still the less well understood. In quantum field theory, electromagnetism and the weak interaction have successfully been embedded into the electroweak interaction \cite{electroweak}, while the strong interaction is very well described by QCD. A common wisdom is that at very high energies, these three interactions combine into a Grand Unified Theory (GUT) based on a gauge group $G_{\rm GUT}$ containing at least the standard model group $SU(3)_{\text{strong}}\times SU(2)_{\text{weak}} \times U(1)_{\text{elec}}$.  As one lowers the energy/temperature, the interactions unified in $G_{\rm GUT}$ start to separate and the standard model emerges. In particular, the electroweak phase transition takes place at low energy  and implies that electromagnetism and the weak interaction split.  While theoretically promising, there is still no experimental evidence for the existence of such a GUT.

Though interesting for studying the last three interactions, the above GUT scenario has an important drawback: It does not include gravity. The oldest known interaction does not  merge with the others in such a consistent quantum field theory. However, it is not currently possible to access experimentally domains of energies relevant for testing a quantum theory of gravity, except maybe through  astrophysical and cosmological observations of phenomena involving very high energies.
Consequently, the realm of quantum gravity has been, and is still, the one for theorists. Over the last thirty years, there has been various attempts to formulate a consistent quantum theory of gravity. Not all of them try to also embed gravity with the other three forces in a  Quantum Theory Of Everything (QTOE). String Theory \cite{String} is a candidate for a QTOE, and it is an important open problem to realize within it not only the standard model of particle physics but also the known cosmology of our Universe.

What we mean by ``known cosmology'' can be described by the Cosmological Standard Model,  dubbed the $\Lambda$-CDM model (Cold Dark Matter). The latter proposes an history of our space-time and predicts its ultimate fate. Starting just after its birth, the Universe underwent a very fast period of acceleration, called \textit{Inflation}. The latter diluted the primordial inhomogeneities, topological relics and rendered the space flat. At the end of inflation, the Universe undergoes a short period of ``reheating" and an era of domination by radiation then appears.  As the temperature lowers, symmetry breaking phase transitions occur (and in particular the electroweak breaking) and the fundamental particles acquire a mass via the Higgs mechanism.  As the Universe cools, hadrons such as the proton and neutron start to form.  Progressively, matter appears as the results of thermonuclear reactions. Consequently, after some time, matter dominates and leads to structure formation. However, it no longer dominates today since there are observational evidences that our Universe is slightly accelerating \cite{Perlmutter}. This is implemented theoretically speaking, by introducing a tiny cosmological constant. For a wider discussion on standard cosmology and the history of our Universe, see \cite{cosmo}.

General Relativity is the theory which best describes gravity in a classical setting.  Supposing the Universe to be 4-dimensional with coordinates $x^{\mu}$ $(\mu=0,\dots,3)$, the Einstein equation takes the form
\be \label{EEq}
G_{\mu\nu} := R_{\mu\nu}-\frac{1}{2}R\, g_{\mu\nu} = T_{\mu\nu},
\ee
in appropriate units such that $c=1$ and $8\pi G=1$. In (\ref{EEq}), $R_{\mu\nu}$ is the Ricci tensor, while $T_{\mu\nu}$ is the stress-energy tensor.
Due to inflation, the $\Lambda$-CDM model treats the Universe as homogeneous and isotropic at sufficiently large scales, and spatially flat. It is then possible to show \cite{FLRW} that the metric describing such a space-time is given by the following FLRW (Friedmann-Lema\^{i}tre-Robertson-Walker) one:
\be \label{FLRW}
ds^2=-N(t)^2dt^2+a(t)^2 \sum_{\mu=1}^{3} (dx^{\mu})^2,
\ee
where $a(t)$ is the \textit{scale factor}. In a homogeneous and isotropic space-time, the stress-energy tensor takes, in the perfect fluid approximation, the form
\be
T_{\mu\nu}=\left(P+\rho\right)u_{\mu}u_{\nu}+Pg_{\mu\nu},
\ee
where $u^{\mu}$ is the $4$-speed of the cosmic fluid, satisfying $u^{\mu}u_{\mu}=-1$. $\rho$ is the energy density and $P$ the pressure.
The $\Lambda$-CDM model describes phenomenologically the observed features of our Universe by coupling Einstein gravity to a matter sector whose field content, potentials and kinetic terms are constrained only by observations.
These sources are described by perfect fluids of species $i=1,2,\dots$ characterized by their densities $\rho_i$ and pressures $P_i$ related by equations of state $P_i=\omega_i\,  \rho_i$, with parameter $\omega_i$.
To be concrete, the $\Lambda$-CDM model states that $97\%$ of the energetic content of the Universe is described by dark matter ($27\%$) and a cosmological constant $\Lambda$ ($70\%$) for dark vacuum energy. While there is much indirect experimental evidence for them, these two quantities still lack direct measurement so that their exact nature is still unknown.  Theoretically speaking, there is a wide diversity of scenarios with both dark matter and cosmological constant candidates (see \cite{BFMS,Zhao:2006vm} for a cosmological scenario trying to explain both dark matter and cosmological constant).
The remaining $3\%$ of the content of the Universe is spanned between pressureless non-relativistic baryonic matter and radiation, the second being a tiny fraction of the first in our present epoch.
Though very interesting, the phenomenological approach of the $\Lambda$-CDM model lacks an underlying microscopic derivation.

It is a challenge for String Theory to provide such a foundation. However, despite considerable efforts toward unraveling string cosmology over the last few years, still very little is known about the dynamics of strings in time-dependent settings. Indeed,  it seems difficult to obtain  time-dependent  solutions in string theory at the classical level. After extensive studies in the framework of superstring compactifications, the obtained results appear to be unsuitable for cosmology. In most cases, the classical ground states correspond to static Anti-de Sitter like or flat backgrounds but not time-dependent ones. The same situation appears to be true in the effective supergravity theories.  Naively, the results obtained in this direction may yield to the conclusion that cosmological backgrounds are unlikely to be found in superstring theory. However, quantum and thermal corrections are neglected in the classical string/supergravity regime. Actually, it turns out that in certain cases, the quantum and thermal corrections are under control  \cite{Cosmo-1}-\cite{Cosmo-4} at the full string level and that cosmological evolutions at finite temperature can be generated dynamically at the quantum level.  The purpose of the present work is to review them and show how some of the weak points of phenomenological approaches can be explored and analyzed concretely in a consistent theoretical framework.  In particular, we will describe how one can find the energy density and pressure from microscopic arguments, by studying the canonical ensemble of a gas of strings.

In order to understand how cosmological solutions arise naturally in this context, we first consider classical supersymmetric flat backgrounds in 4 dimensions.  They are obtained from the $10$-dimensional space-time in which superstrings are living by compactifying 6 directions.
The study of the thermodynamics of a gas of superstring states filling this background makes sense at the quantum level only (this is well known from Planck, when he introduced the notion of quanta to solve the UV catastrophe problem in black body physics). At finite temperature, the quantum and thermal fluctuations
produce a non-zero free energy density which is computable perturbatively at the full string level.
Note that in this context, it  can be determined order by order in the Riemann surface genus expansion without the UV ambiguities encountered in the analogous computation in quantum field theory. An energy density and pressure can be derived from the free energy. Their back-reaction on the space-time metric and moduli fields (the continuous parameters of the models) gives rise to specific cosmological evolutions.
In this review, the above strategy is restricted to the domain of temperatures lower than the Hagedorn temperature and higher than the electroweak breaking scale to be specified in the next paragraphs. In this intermediate regime,  the evolution of the Universe is found to converge to a radiation dominated era.

More interesting models are those where space-time supersymmetry is spontaneously broken at a scale $M$ before finite temperature is switched on.  With the supersymmetry breaking mechanism we consider, the stringy quantum corrections are under control in a way similar to the thermal ones \cite{Cosmo-1} --\cite{Cosmo-4}. In large classes of models, the back-reaction
of the quantum and thermal corrections on the space-time metric and the moduli fields
induces a cosmological evolution which is attracted to a radiation dominated era. The latter is characterized by a temperature and a supersymmetry breaking scale that evolve proportionally to the inverse of the scale factor, $T(t)\propto M(t)\propto 1/a(t)$.

In the context of string theory, we can study much higher energies than the ones tested so far. We are then limited by the appearance of a \textit{Hagedorn phase transition} at ultra high temperature \cite{Hagedorn}. It is a consequence of the exponential growth of the number of states that can be thermalized at high temperature and implies a divergence of the canonical thermal partition function.
The latter can be computed in Matsubara formalism \ie in Euclidean time compactified on an circle of circumference $\beta$, the inverse temperature.  The Hagedorn instability is signaled by string modes wrapping the Euclidean time circle that become tachyonic (\ie with negative (mass)$^2$) when the temperature is above the Hagedorn temperature $T_H$ \cite{AtickWitten,RostKounnas}.  The possible existence of an Hagedorn era in the very early Universe provides a possible alternative or at least a complementary point of view to inflation, as developed in \cite{BDB}. However, we do not discuss this very high temperature regime here,  and we will consider the physics at low enough temperatures compared to $T_H$ to avoid the occurrence of a Hagedorn phase transition \cite{AKADK}.
Note that models free of Hagedorn instabilities \cite{Carlo,MSDS} and still under computational control can also be constructed.
In the present review, we bypass the Hagedorn era ambiguities by assuming that 3 large spatial directions have emerged before $t_E$ (the exit time of the Hagedorn era), along with internal space directions whose size characterizes the scale of spontaneous supersymmetry breaking. Within this assumption, we parameterize our ignorance of the detailed physics in the Hagedorn era by considering arbitrary initial boundary conditions (IBC) for the fields at $t_E$.

At late cosmological times, when the temperature of the Universe is low enough,  it is possible for an additional scale $Q$ to become relevant. $Q$ is the
infrared renormalisation group invariant transmutation scale induced at the quantum level by
the radiative corrections of the soft supersymmetry breaking terms at low energies \cite{Q-Noscale}. When
$T(t) \sim Q$, the electroweak phase transition takes place, $SU(2) \times U(1) \rightarrow U(1)_{elec}$. This starts
to be the case at a time $t_W$ and, for $t > t_W$, the supersymmetry breaking scale $M$
is stabilized at a value close to $Q$. In earlier cosmological times where $M(t) \sim T(t) > Q$, the
transmutation scale $Q$ is irrelevant and the Universe is in the radiation era. It turns out that the electroweak symmetry breaking transition is very sensitive to the specifics of the string background considered, while in the earlier radiation era the results are fairly robust.
We restrict our analysis to the intermediate cosmological times:
\be
 t_E \ll t \ll t_W,
\ee
namely, after the exit of the Hagedorn era and before the electroweak symmetry breaking.

In section 2, we present the basics of our approach and apply it to the simplest examples where supersymmetry is spontaneously broken by temperature effects only. In this class of models, the moduli (radii) of the internal space are held fixed, close to the string scale. In section 3, we analyze models where supersymmetry is spontaneously broken even at zero temperature. We take into account the dynamics of the supersymmetry breaking scale $M$ which is  a field and keep frozen the other moduli. This is only in section 4 that we show the  latter hypothesis is consistent by taking into account the dynamics of internal radii that are not participating in the breaking of supersymmetry. The last section is devoted to our conclusions.

%%%%%%%%%%%%%%%%%%%%%%%%%%%%%%%%%%%%%%%%%%%%
%%%%%%%%%%%%%%%%%%%%%%%%%%%%%%%%%%%%%%%%%%%%

\section{Basics of our approach}

\subsection{Thermodynamics and variational principle}

Let us first describe how thermodynamical results can be derived from general relativity. As a simple example, we consider the gas of a single bosonic state at temperature $T$ in a  3-dimensional torus $T^3$, which is nothing but a box with periodic boundary conditions and large volume $V_{\rm box}=(2\pi R_{\rm box})^3$. The number of particles is not fixed and the canonical ensemble partition function $Z_{\rm th}$ is defined in terms of the Hamiltonian $H$ and inverse temperature $\beta$. In second quantized formalism, $Z_{\rm th}$ can be expressed as a path integral
\be
\label{Zther}
Z_{th}:={\rm Tr\, } e^{-\beta H}=\int \mathcal{D}\varphi \, e^{-S_E[\varphi]},
\ee
where $S_E$ is the Euclidean action of the quantum field $\varphi$ and the Euclidean time is compact with period $\beta$. The boundary condition along the Euclidean circle of the bosonic field $\varphi$ is periodic (while a fermionic field would be anti-periodic). The thermal partition function can be written in terms of an infinite sum of connected or disconnected Feynmann graphs. Supposing the gas to be (almost) perfect, the particles do not interact (much) with themselves and we can approximate the result at one loop. The free energy takes the form
\be
\label{Fth}
F=-{\ln Z_{\rm th}\over \beta} \simeq -{Z_{1-\rm loop}\over \beta},
\ee
where $Z_{1-\rm loop}$ is the unique connected graph at 1-loop, the \textit{bubble diagram}, a single propagator whose two ends are identified. Then, one can derive from $F$ the energy density and pressure using standard thermodynamics identities.

However, it is also possible to use the fact that $Z_{1-\rm loop}$ is, from a quantum field theory point of view, the 1-loop vacuum-to-vacuum amplitude  \ie vacuum energy inside the box. Viewing the whole space we are living in as the box itself, the classical Einstein action must be corrected at 1-loop by a contribution to the ``cosmological constant'',
\be
\label{RG}
S = \int d^4x \sqrt{-g} \, \left( {R\over 2}+{Z_{\rm 1-loop}\over \beta V_{\rm box}}\right).
\ee
Note that in the above action, we are back to real time \ie {\em Lorentzian} signature, by analytic continuation on the time variable. The stress-energy tensor is found by varying with respect to the metric,
\be
T_{\mu\nu}= -{2\over \sqrt{-g}}\, {\delta\over \delta g^{\mu\nu}}\left(\sqrt{-g}\; {Z_{\rm 1-loop}\over \beta V_{\rm box}}\right).
\ee
Originally, the classical background is homogeneous and locally flat Minkowski space. Its metric is of the form (\ref{FLRW}) with laps function $N=\beta$ and scale factor $a=2\pi R_{\rm box}$, as follows from the analytic continuation of the Euclidean background in which the 1-loop vacuum-to-vacuum amplitude has been computed. With $Z_{\rm 1-loop}$ a function of $\beta$ and $V_{\rm box}$, the stress-energy tensor  takes the form $T^\mu {}_\nu = \text{diag}(-\rho,P,P,P)^\mu{}_\nu$, where
\begin{eqnarray}
\label{pressen}
&P\!\!\!&={1\over \beta}\, {\partial Z_{\rm 1-loop}\over \partial V_{\rm box}}\equiv -\left(\frac{\partial F}{\partial V_{\rm box}}\right)_\beta\\
&\rho\!\!\!&=-{1\over V_{\rm box}}\, {\partial Z_{\rm 1-loop}\over \partial \beta}\equiv {1\over V_{\rm box}}\left({\partial(\beta F)\over \partial \beta}\right)_{V_{\rm box}}.
\end{eqnarray}
In these expressions, the right hand sides follow from Eq. (\ref{Fth}) and reproduce the standard thermodynamical results. When $Z_{\rm 1-loop}$ is proportional to $V_{\rm box}$ \ie the free energy is extensive, these relations simplify to
\be
P=-{\F}\qquad \and \qquad \rho=T\, {\partial P\over \partial T}-P,
\ee
where $\F:= \displaystyle{F\over V_{\rm box}}$ is the free energy density and the relation between $\rho$ and $P$ is the state equation.

The above approach has the advantage to allow to go farther than deducing the thermodynamical identities. As long as the 1-loop sources $\rho$ and $P$ are small perturbations of the classically homogeneous and isotropic static background, one can find their back-reaction on the space-time metric by solving the Einstein equation (\ref{EEq}). In other words, a quasi-static evolution $\beta(t)$, $a(t)$ is found. This hypothesis amounts to supposing that the evolution is a sequence of thermodynamical equilibria \ie that it is slow enough for the temperature to be remain homogenous.

We could have been satisfied by this quantum field theory approach in general relativity if an important difficulty would not arise: In most cases, $Z_{\rm 1-loop}$ is actually divergent ! This is always the case for a single bosonic (or fermionic) field. However, suppose the gas contains two species of free particles, one bosonic of mass $M_B$ and one fermionic of mass $M_F$. The 1-loop vacuum-to-vacuum amplitude is found to be
\be
\label{Zft}
Z_{\rm 1-loop}=\beta V_{\rm box}\int_0^{+\infty}{dl\over 2l}\, {1\over (2\pi l)^2}\sum_{\tilde m_0}\left(e^{-{ l\over 2}M^2_B}-(-)^{\tilde m_0} e^{-{ l\over 2}M^2_F}\right) e^{-{\beta^2\tilde m_0^2\over 2l}},
\ee
where $l$ is a ``Schwinger parameter'', the proper time of each particle when it runs into a loop wrapped $\tilde m_0$ along the Euclidean time circle. If $M_B\neq M_F$ (that could arise from a spontaneous breaking of supersymmetry), the contribution to this integral with $\tilde m_0=0$ is divergent in the UV, $l\to 0$. We conclude that only the exactly supersymmetric spectrum $M_B=M_F$ gives a well defined free energy.

As is well know, the amplitudes in string theory are free of UV divergences and one can expect that the above approach applied in this context will give a perfectly well established framework to describe thermodynamics in time-dependent backgrounds. Actually, we are going to see that  string theory provides a rigorous microscopic derivation of the sources $\rho$ and $P$ and their equation of state.

%%%%%%%%%%%%%%%%%%%%%%%%%%%%%%%%%%%%%%%%%%%%
%%%%%%%%%%%%%%%%%%%%%%%%%%%%%%%%%%%%%%%%%%%%

\subsection{Supersymmetric string models at finite \em T}

We start to implement the ideas sketched at the end of the previous section on simple string theory models in 4-dimensional Minkowski space-time, where supersymmetry is spontaneously broken by thermal effects only. The case of models where supersymmetry is spontaneously broken even at zero temperature is addressed in the next section.

To be specific, we consider heterotic models but type II or type I ones can be treated similarly. To analyze the canonical ensemble of a gas of heterotic strings, we consider 10-dimensional backgrounds of the form
\be
\label{back}
\underbrace{S^1(R_0)}_{\text{Euclidean time}} \times\;\; \;\underbrace{T^{3}(R_{\rm box})}_{\text{space}}\;\; \;\times \underbrace{{\cal M}_{6}}_{\text{internal space}},
\ee
where $S^1(R_0)$ is the Euclidean time circle of perimeter $\beta=2\pi R_0$, $T^3(R_{\rm box})$ denotes the spatial part which is taken to be flat and compact \ie a very large torus, and ${\cal M}_{6}$ represents the remaining internal manifold.

Classically, the vacuum energy vanishes. This is clear from the fact that the genus-0 vacuum-to-vacuum string amplitude is computed on the Riemann sphere, which is simply connected and thus cannot wrap the Euclidean time. Consequently, it cannot probe the temperature effects that are responsible of the breaking of supersymmetry. However, the genus-1 Riemann surface which is nothing but a torus has two cycles that can wrap the Euclidean time so that a non-trivial 1-loop contribution to the vacuum-to-vacuum energy arises, $Z_{\rm 1-loop}$. To be specific, we focus on the simplest example where ${\cal M}_6=T^6$, which means that at $T=0$ the model is $\N=4$ supersymmetric in 4 dimensions for the heterotic case.
Using world-sheet techniques, $Z_{\rm 1-loop}$ is given by \cite{Cosmo-1}-\cite{Cosmo-4}
\be
\label{partAp}
Z_{\rm 1-loop} ={\beta V_{\rm box}\over (2\pi)^4} \int_{\mathcal{F}} {d \tau d \bar \tau \over 4 (\Im \, \tau)^{3}}\,  {1 \over 2} \sum_{a,b} (-1)^{a+b+a b}\,  \frac{\vartheta^4 \ab}{\eta^4} \, {\Gamma_{(6,22)} \over \eta^{8} \bar \eta^{24}} \sum_{\tilde m_0,n_0}e^{-{\pi R_0^2\over \Im \, \tau}|\tilde m_0+n_0\tau|^2} (-1)^{\tilde m_0 a + \tilde n_0 b + \tilde m_0 n_0}.
\ee
A few words might be helpful to understand this expression. On the 2-dimensional world-sheet of the heterotic string, there are left moving superstring waves and right moving bosonic ones. There are thus $10$ left moving world-sheet fermions and bosons, and $26$ right moving world-sheet bosons. However, ghosts cancel the contributions of two bosons (left and right) and two left fermions. The remaining fermions contribute the factor $\vartheta^4 \ab(\tau) /\eta^4(\tau)$ with an appropriate spin-statistic phase $(-1)^{a+b+a b}$ depending on the integers $a,b$ modulo 2 associated to the boundary conditions of the fermions along the two circles of the world-sheet torus. The contributions of the left and right moving world-sheet bosons correspond to the $\Gamma_{(6,22)}(\tau,\bar\tau)/\eta^8(\tau)/\bar \eta^{24}(\bar\tau)$ factor. $\Gamma_{(6,22)}$ is a lattice that corresponds to the 0-modes of the  $16$ right moving bosons (without left partners) and the six internal left and right moving bosons that realize the coordinates of the internal space $T^6$.
In field theory, the loop can wrap $\tm_0$ times around the Euclidean time. In string theory, the closed string itself can also be wrapped $n_0$ times around $S^1(R_0)$. This is why we have a double discrete sum on arbitrary integers $\tm_0,n_0$ and a generalized phase $(-1)^{\tilde m_0 a + \tilde n_0 b + \tilde m_0 n_0}$ that involves the winding number $n_0$, as compared to Eq. (\ref{Zft}), where space-time bosons (fermions) correspond to $a=0$ ($a=1$).

Another important difference between the field and  string theory amplitudes, Eqs (\ref{Zft}) and  (\ref{partAp}),  is that the integration over the Schwinger parameter from 0 to $+\infty$ is replaced by an integral over the {\em fundamental domain} of $SL(2,\mathbb{Z})$,
\be
\mathcal{F}= \left\{\tau \in \mathbb{C}\;\;  \Big/\;\;  |\Re \, \tau |\leq \frac{1}{2},\; \;  \Im \, \tau>0, \; \;  |\tau|\ge 1\right\},
\ee
which does not contain the line $\Im \, \tau =0$. Since the ``Schwinger parameter'' or proper time along the string world-sheet torus is $\Im \, \tau$, there is no risk of any UV divergence. This property of the amplitude is only due to the extended nature of the string that provides a natural cut-off in the UV. Contrarily to the field theory case, the UV finiteness of the amplitude is guaranteed in all models, even when the spectrum at zero temperature is not supersymmetric (see section 3). However, both in field and string theory, the amplitude is finite in the IR ($l$ and $\Im \, \tau \to +\infty$) as long as there is no tachyon in the spectrum (\ie no particle with (mass)$^2<0$). A careful analysis shows that string states winding around the Euclidean time circle become tachyonic when $1/R_H<R_0<R_H$, where $R_H=\displaystyle {1+\sqrt{2}\over 2}$ in $\sqrt{\alpha'}$ units \cite{AtickWitten} --\cite{AKADK}, the string length we have set to 1 in Eq. (\ref{partAp}) and now on for notational simplicity. $R_H$ determines the Hagedorn temperature at which a phase transition occurs. As announced in section 1, we restrict ourselves to the study of epochs in the history of the Universe that follow the Hagedorn era and thus consider only regimes where $R_0\gg 1$.

In the amplitude (\ref{partAp}) (or  (\ref{Zft}) in field theory), the dominant contributions arise from the lightest states. The pure Kaluza-Klein (KK) states associated to the Euclidean time circle have masses of order $1/R_0$. Strings with non-trivial winding number $n_0$ around $S^1(R_0)$ get a contribution to their mass proportional to the length of the circle \ie $R_0$. Similarly, the KK and winding states associated to the internal space ${\cal M}_6$ have masses contributions of order $1/R_I$ and $R_I$, where $R_I$ denotes some generic radius (modulus) characterizing the size of ${\cal M}_6$. Finally, each string that oscillates has a contribution of order 1 to its mass. For simplicity in this section, we suppose that all radii $R_I$ are satisfying the constraint
\be
\label{inter}
{1\over R_0}\ll R_I \ll R_0
\ee
that will be justified in section 4.\footnote{It will be shown that for arbitrary I.B.C. at the exit of the Hagedorn era, $R_I$ is dynamically attracted to the interval $1/R_0<R_I<R_0$ and then converges to a constant.}  It follows that the towers of pure KK states along the Euclidean time are much lighter than any other states in the spectrum. Given that, the partition function (\ref{partAp}) can be written as
\be
\label{Zstef}
Z_{\rm 1-loop}={\beta V_{\rm box}}\, {1\over (2\pi R_0)^4}\, n_T \, c_4+\cdots \qquad \text{ where } \qquad c_4={1\over \pi^{2}}\sum_{\tm_0}{1\over \abs 2\tm_0+1\abs^4}={\pi^2\over 48}
\ee
and the dots stand for contributions of order $\O(e^{-2\pi R_0})$ for the oscillating states, $\O(e^{-2\pi R_0/R_I})$ and $\O(e^{-2\pi R_0R_I})$ for the KK and the winding states of ${\cal M}_6$, and $\O(e^{-2\pi R^2_0})$ for winding states around $S^1(R_0)$. These terms are all exponentially suppressed, compared to the dominant contribution. In Eq. (\ref{Zstef}),  $n_T$ is the number of massless boson-fermion pairs in the supersymmetric model when the temperature is not switched on. The constant $c_4$ is a dressing that accounts for the full towers of KK states along   $S^1(R_0)$.

Let us determine the back-reaction of the non-trivial 1-loop vacuum energy on the originally static background. At order one in string perturbation theory, the low energy effective action at finite temperature is
\be
\label{Sth}
S = \int d^4x \sqrt{-g_{st}} \left[e^{-2\phi}\left({R_{st}\over 2}+2(\partial \phi)^2\right)+{Z_{\rm 1-loop}\over \beta V_{box}}\right],
\ee
where $\phi$ is the dilaton in four dimensions. Compared to the general relativity case Eq. (\ref{RG}), the string ``coupling constant'' is actually the field $e^{2\phi}$. The choice in the definition of the metric tensor that gives rise to the above mixing of the dilaton and the Ricci curvature is referred as the string frame metric.
In (\ref{Sth}), we have kept constant the over massless fields of the  string spectrum since the 1-loop source does not involve them in the present case (see the next sections for more general models). Their kinetic terms are thus vanishing.

The action may be converted to a more convenient ``Einstein frame'' by rescaling the metric as $g_{st \mu\nu} = e^{2 \phi}\,  g_{\mu\nu}$,
\be
\label{SthE}
S=\int d^4x \sqrt{-g} \left[{R\over 2}-(\partial \phi)^2-\F \right]\, ,
\ee
where
\be
\F=-T^4\,  n_T\,  c_4\qquad \and \qquad T={1\over 2\pi R_0\, e^{-\phi}}.
\ee
$\F$ and $T$ which contain a dilaton dressing are the free energy density and temperature when they are measured in Einstein frame. Supposing that the back-reaction of the thermal sources induce a quasi-static evolution of the homogeneous and isotropic background, dilaton and temperature, we consider an ansatz
\be
\label{ansatz}
\begin{array}{l}
ds^2=-N(x^0)^2(dx^0)^2+a(x^0)^2\left[(dx^1)^2+(dx^2)^2+ (dx^{3})^2\right]\; , \qquad \phi(x^0)\, , \\
\where \displaystyle\qquad N(x^0)\equiv 2\pi R_0 \, e^{-\phi}\equiv {1\over T(x^0)}\; ,\qquad a(x^0)\equiv 2\pi R_{box} \, e^{-\phi}\, .
\end{array}
\ee
The derivation of the stress-energy tensor reaches
\be
\rho=3P\qquad \where \qquad P=T^4\,  n_T\,  c_4,
\ee
which is nothing but Stefan's law for radiation. This was expected since under the hypothesis (\ref{inter}), all non-zero masses are of order 1 which is also the scale of $T_H$. Since we consider temperatures far below the Hagedorn one, only the massless states can be thermalized. The massive ones remain ``cold'' \ie decoupled from the thermal system.

The equations of motion are easily solved. In terms of cosmological time such that $Ndx^0=dt$, the velocity of the dilaton, $\dot \phi\propto 1/a^3$, goes to zero when the Universe expands. The evolution is thus attracted to the particular solution where $\phi$ is constant \ie the cosmology of a Universe filled by a radiation fluid:
\be
a(t)=\sqrt{t}\times a_0T_0(n_Tc_4)^{1/4}={1\over T(t)}\times a_0T_0\; , \qquad \phi=cst.,
\ee
where $a_0$, $T_0$ are integration constants.
We shall refer to such an attractor as a Radiation Dominated Solution (RDS).

Though we started with the huge machinery of string theory, we finally ended with standard results when the Universe is filled with radiation, after we consider the dominant contribution of massless modes. The reader could be skeptical about the need to require such heavy tools to describe such simple physics. However, we remind that our approach gives a rigorous microscopic derivation of these results that will be generalized in sections 3 and 4 to models whose free energies are UV divergent in field theory. In addition, we are going to see that when the dynamics of other scalar fields is taken into account, the underlying string theory provides a connection between naively disconnected theories as seen from a field theory point of view. This string theoretic effect marks the novelty of our approach.

%%%%%%%%%%%%%%%%%%%%%%%%%%%%%%%%%%%%%%%%%%%%
%%%%%%%%%%%%%%%%%%%%%%%%%%%%%%%%%%%%%%%%%%%%

\section{Non-supersymmetric string models at finite \em T}

The aim of the previous section was to present the basic ideas of a string theory framework able to provide a microscopic origin for source terms for the gravitational (and moduli) fields. We considered a model $\N=4$ supersymmetric when temperature is not switched on. In order to recover a non-supersymmetric physics at very low temperature \ie late time from a cosmological point of view, we need to consider models whose spectra are not supersymmetric, even at $T=0$. From a phenomenological point of view, we are particularly interested in models with $\N=1$ spontaneously broken  supersymmetry.

To switch on finite temperature, we have introduced periodic or antiperiodic boundary conditions on  the Euclidean time circle for the string states, depending on their fermionic number ($a=0$ for bosons and $a=1$ for fermions in Eq. (\ref{partAp})). In a similar way, a spontaneous breaking of supersymmetry can be generated by non-trivial boundary conditions on internal circles $S^1(R_i)$ ($i=4,\dots, 3+n$), using R-symmetry charges $a+Q_i$. Both finite temperature and supersymmetry breaking implemented this way can be thought as a string theoretic generalizations of Scherk-Schwarz compactifications \cite{SS}. Physically speaking, this can be thought as introducing non-trivial background fluxes along the cycles. Two mass scales
then appear and are {\it a priori} time-dependent: The temperature $T \propto \frac{1}{2\pi R_0}$ and the supersymmetry breaking scale $M \propto \frac{1}{2\pi (\prod_i R_i)^{1/n}}$. The initially degenerate mass levels of bosons and fermions split by amounts proportional to $T$ and/or $M$. This mass splitting is the signal of supersymmetry
breaking and gives rise to a non-trivial free energy density at 1-loop. Note that in the pure thermal case, each originally degenerate boson-fermion pair gives a positive contribution to $Z_{\rm 1-loop}$ since it is always the fermion that is getting a mass shift (their momenta along $S^1(R_0)$ are half integer).
However, when introducing supersymmetry breaking, bosons can have non-trivial R-symmetry charges and acquire masses bigger than the ones for fermions. Consequently, negative contributions  to the vacuum energy can arise.
This will play an important role for RDS to exist. In the following, we first consider simple examples of models with $n=1$ before sketching some cases with $n=2$ to observe how ratios of radii (commonly referred as  ``complex structure moduli'') can be stabilized.

%%%%%%%%%%%%%%%%%%%%%%%%%%%%%%%%%%%%%%%%%%%%

\subsection{Supersymmetry breaking involving {\em n} = 1 internal dimension}
\label{n=1mo}

We want to study the canonical ensemble of a gas of heterotic strings, where supersymmetry is spontaneously broken by non-trivial boundary conditions along the internal direction 4. To be specific, we focus on two kinds of backgrounds (\ref{back}), with internal space
\be
\label{I,II}
({\rm I})\; : \quad  \M_6=S^1(R_4)\times S^1 \times \frac{T^4}{\mathbb{Z}_2}\qquad \quad \mbox{or}\quad \qquad ({\rm II})\; : \quad \M_6=\frac{S^1(R_4)\times T^3}{\mathbb{Z}_2} \times T^2.
\ee
$\Z_2$ acts as $x^I\to -x^I$, where $I=6,7,8,9$ in case (I) and $I=4,5,6,7$ in case (II), and breaks explicitly half of the supersymmetries. Thus, the above models have spectra where $\N=2$ supersymmetry is spontaneously broken to zero by the internal flux around $S^1(R_4)$ and the temperature effects. Models with $\N=1\to 0$ can also be analyzed by considering $\displaystyle {\cal M}_6={S^1(R_4)\times T^5\over \Z_2 \times \Z_2}$. They share similar cosmological properties with the models in case (II).

As in the pure thermal case of section 2, tachyonic instabilities arise when $R_0$ or $R_4$ approach $R_H$. In order for the canonical ensemble to be well defined, we will restrict our analysis to regimes where $R_0\gg 1$ and $R_4\gg 1$. We also suppose that all internal radii that are not participating in the spontaneous breaking of supersymmetry satisfy
\be
\label{inter2}
{1\over R_0}\ll R_I\ll R_0\qquad \and \qquad {1\over R_4}\ll R_I\ll R_4 \qquad (I\neq 4),
\ee
and remind the reader that section 4 is devoted to the justification that this is consistent with the dynamics of the $R_I$'s. Under these hypothesis, the 1-loop partition function of the pure thermal case (\ref{Zstef}) is generalized to
\be
\label{Z2}
Z_{\rm 1-loop}=\beta V_{\rm box}\left( {1\over (2\pi R_0)^4}\, n_T\, \hat f^{(4)}_T(z)+ {1\over (2\pi R_4)^4}\, n_V\, e^{-z}\hat f^{(4)}_T(-z)+ {1\over (2\pi R_0)^4}\,n'_T\, c_4+\cdots\right),
\ee
where
\be
\label{ZTM}
\hat f_T^{(4)}(z)=\frac{\Gamma(5/2)}{\pi^{5/2}}\sum_{\tk_0,\tk_4 \in \mathbb{Z}}\frac{e^{4z}}{[(2\tk_0+1)^2e^{2z}+(2\tk_4)^2]^{5/2}}\; , \qquad e^z={R_0\over R_4}.
\ee
In Eq. (\ref{Z2}), $R_0$ and $R_4$ being large, the modes which are KK excitations along both $S^1(R_0)$ and $S^1(R_4)$ give dominant contributions corresponding to the two first terms in the parenthesis. In other words, while $R_i/R_I$ and $R_iR_I$ ($i=0,4$ and $I\neq 4$) are very large and give exponentially suppressed contributions we neglect, it is not necessary the case for $R_0/R_4$. As a consequence, the dimension full factors $1/R_0^4$ and $1/R_4^4$  are dressed with non-trivial functions of the ``complex structure'' ratio $e^z=R_0/R_4$.
In case (I), $n_T$ is the number of massless states (before we switch on finite temperature). $n_V$ is the number of massless bosons minus the number of massless fermions (before we switch on finite temperature) and depends on the choice of R-symmetry charge $a+Q_4$ used to break spontaneously supersymmetry. Both $n_T$ and $n_V$ contain contributions arising from the untwisted and twisted sectors of the $\Z_2$-orbifold, while $n_T'=0$.
In case (II), $n_T$  and $n_V$ are defined similarly, but contain contributions from the untwisted sector only. On the contrary, the mass spectrum in  the $\Z_2$-twisted sector does not depend on $R_4$ and supersymmetry between the associated states is spontaneously broken by thermal effects only. $n_T'$ is then the number of massless boson/fermion pairs in the twisted sector (before we switch on finite temperature).
To summarize, we have
\be
n_T>0\; ,\qquad    -1\le {n_V\over n_T}\le 1    \; ,\qquad n_T'=0\;  \mbox{ in case (I)}\; ,  \;  \; \; \quad n_T'>0 \; \mbox{ in case (II)}.
\ee

As before, $Z_{\rm 1-loop}$ backreacts on the classical 4-dimensional Lorentzian
background, via the effective field theory action,
 \be
 \label{action}
S = \int d^4x \sqrt{-g_{st}}\left[ e^{-2\phi}\left({R_{st}\over 2}+2(\partial \phi)^2+
 {1\over2} (\partial \ln R_4)^2\right)+{Z_{\rm 1-loop}\over \beta V_{\rm box}}\right] ,
\ee
where the scalar kinetic term of $R_4$ is included since there is a non-trivial source at 1-loop for it.
Before writing the equations of motion, it is useful to redefine the scalar fields as,
\be
\Phi:= \sqrt{2\over 3}\, (\phi-\ln R_4)\; , \qquad \phi_\bot := {1\over
  \sqrt{3}}\, (2\phi+\ln R_4)\, ,
\ee
and switch from string to Einstein frame metric. The action becomes
 \be
S = \int d^4x \sqrt{-g}\left[ {R\over 2}-{1\over 2}\left((\partial
    \phi)^2+(\partial \phi_\bot )^2\right)-\F \right] ,
\ee
where  $\F$ is the free energy density,
\be
\F=-T^4\left(n_T\, \hat f^{(4)}_T(z)+n_V\, e^{3z} \hat f^{(4)}_T(-z)+n_T'\, c_4\right):=-T^4\, p(z),
\ee
and $z$ can be expressed in terms of the temperature and supersymmetry breaking scales as,
\be
e^z={M\over T}\; , \qquad T={1\over 2\pi\,  R_0 e^{-\phi}}\; , \qquad M={1\over 2\pi R_4 \, e^{-\phi}}\equiv {e^{\sqrt{3\over 2}\, \Phi}\over 2\pi}.
\ee

\begin{figure}[h!]
\begin{center}
\vspace{.3cm}
\includegraphics[height=3.5cm]{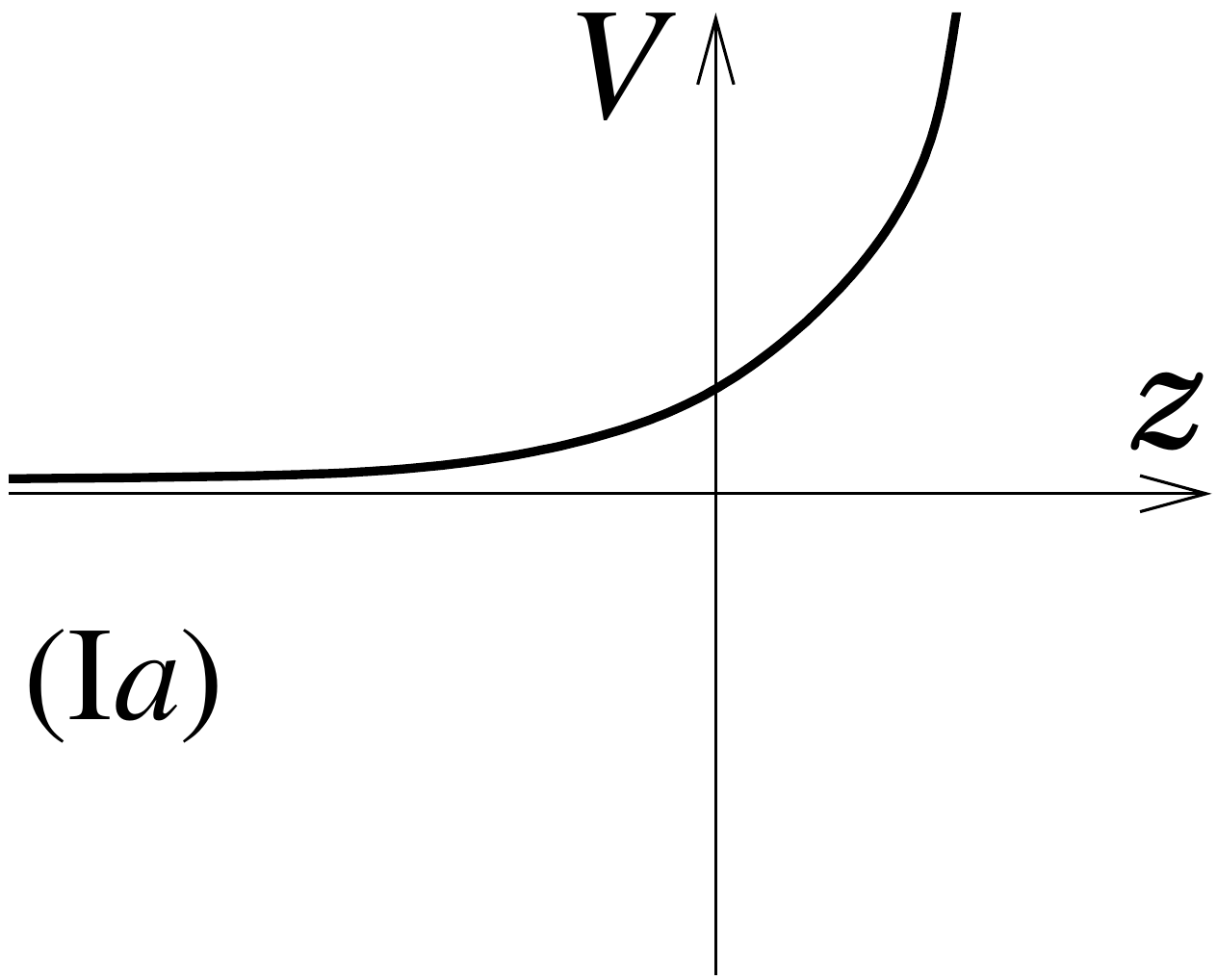}\qquad\quad
\includegraphics[height=3.5cm]{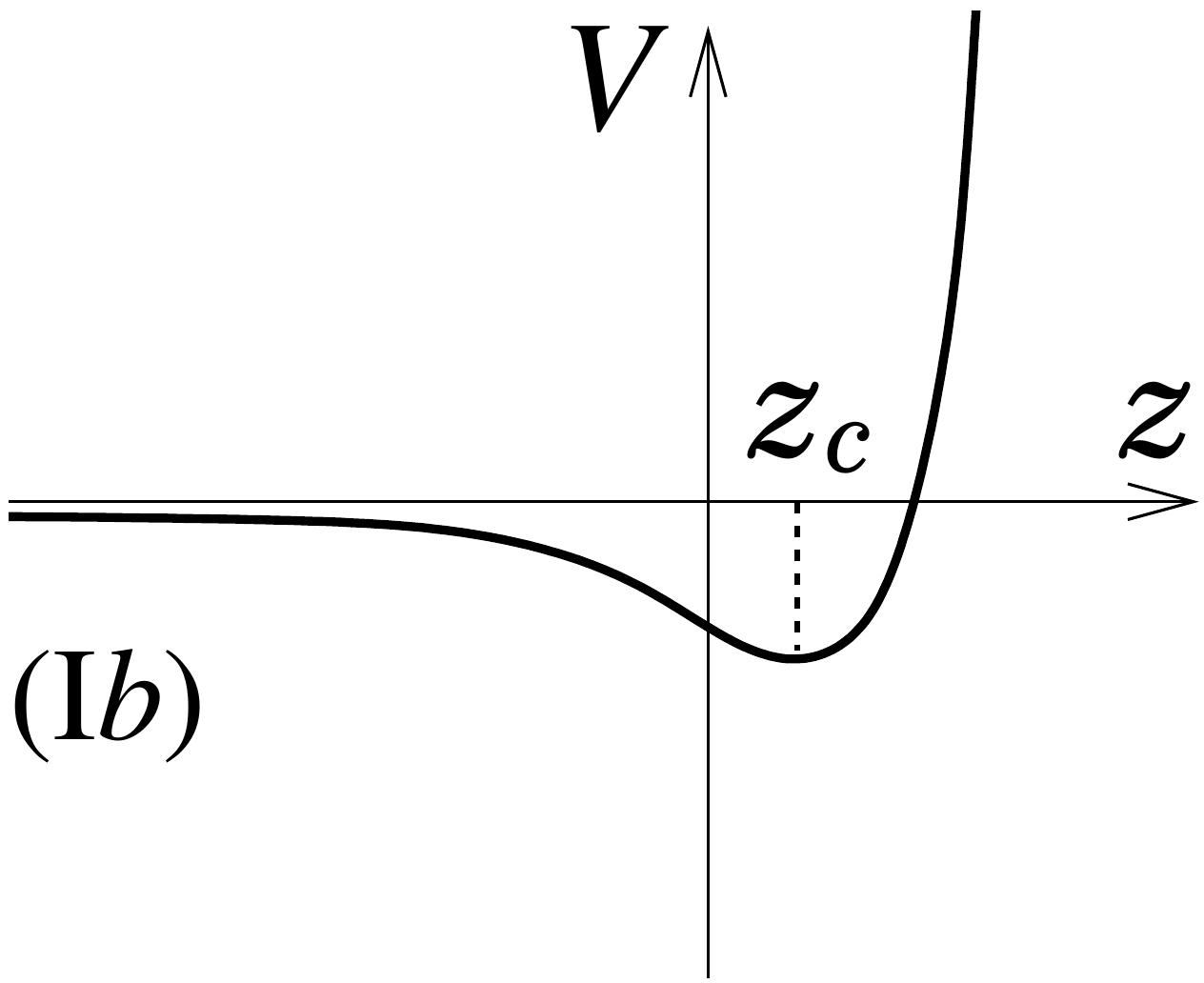}\qquad\quad
\includegraphics[height=3.5cm]{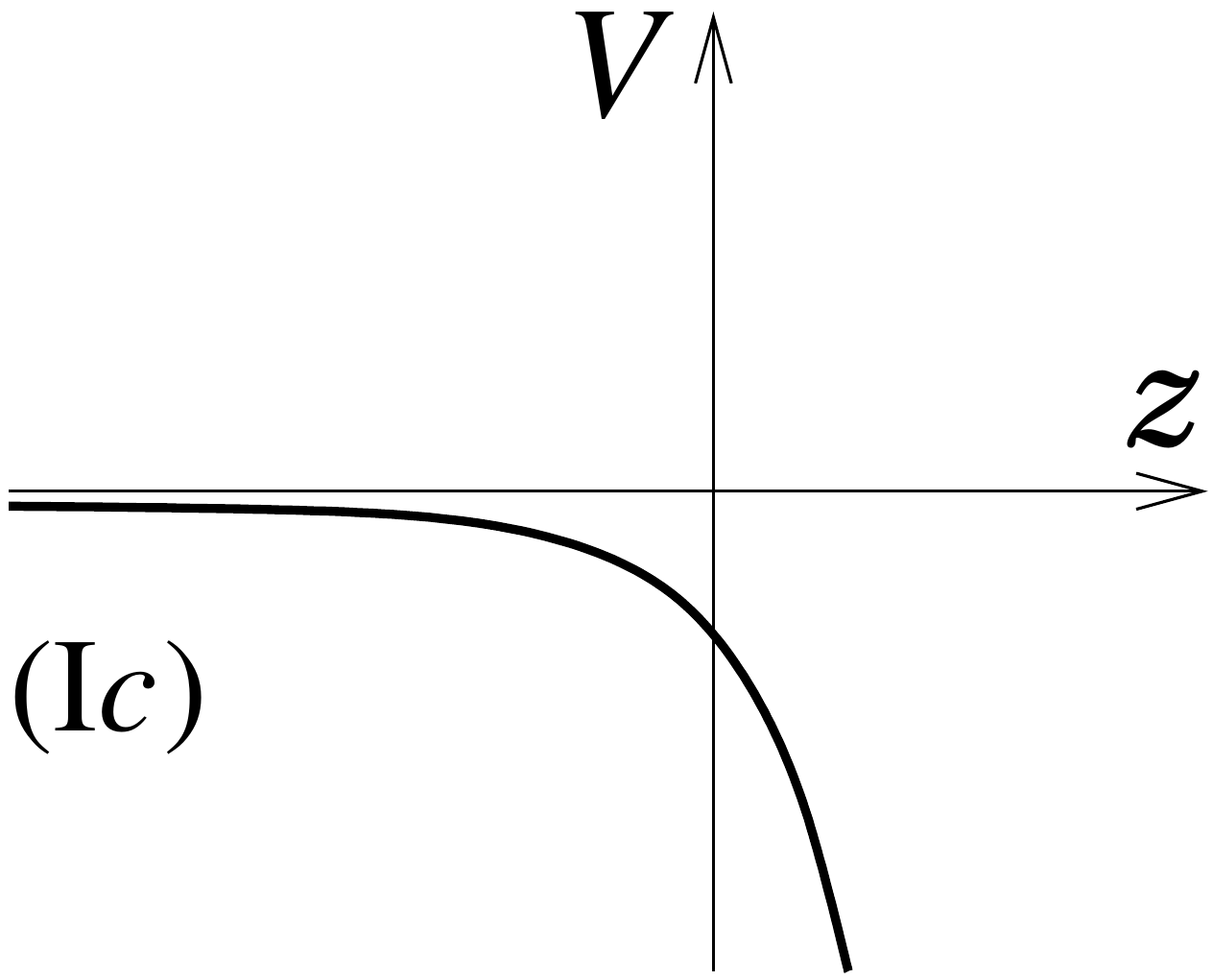}\\
\vspace{.4cm}
\includegraphics[height=3.5cm]{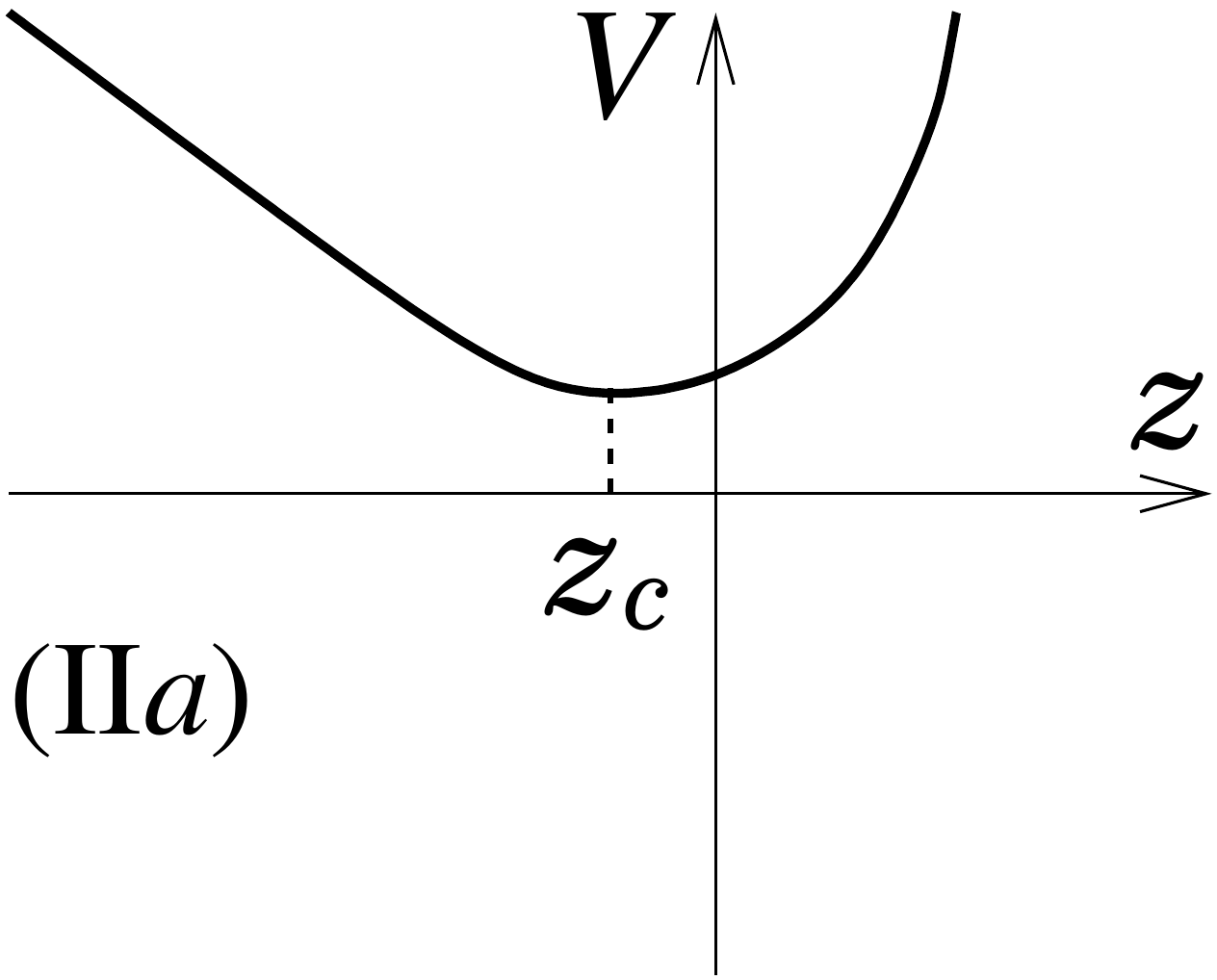}\qquad \quad
\includegraphics[height=3.5cm]{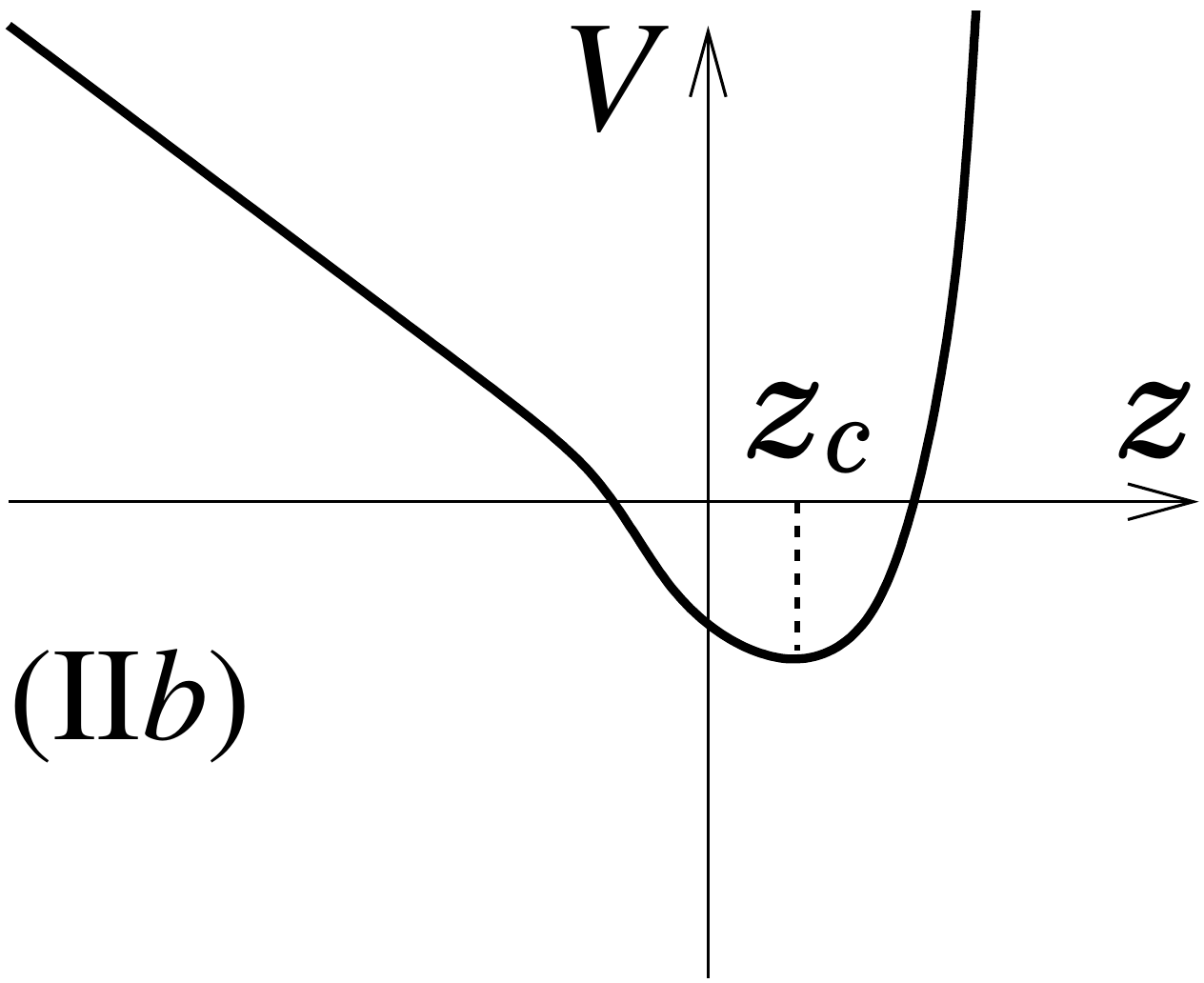}\qquad\quad
\includegraphics[height=3.5cm]{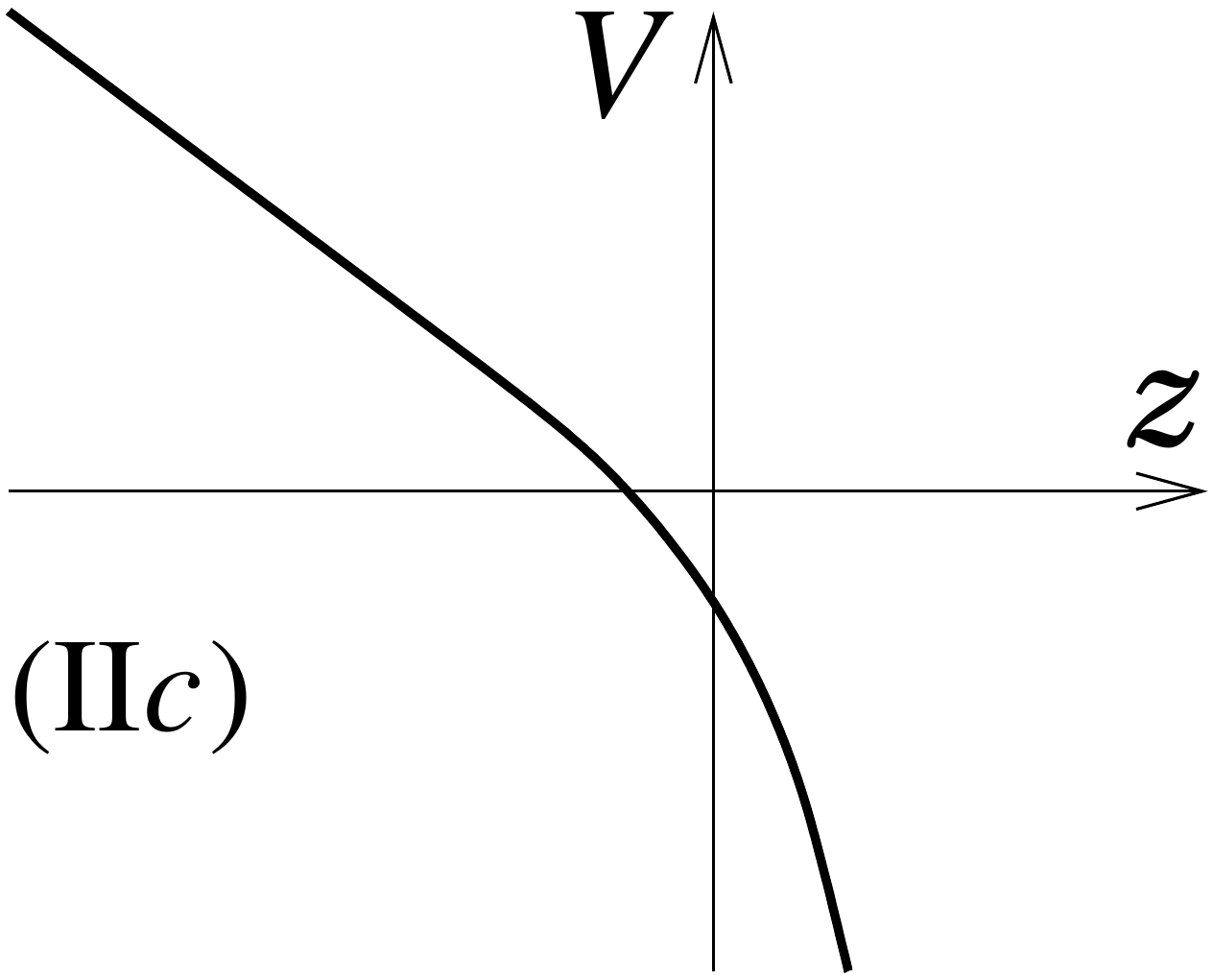}
\caption{\footnotesize \em Different qualitative behaviors of $V(z)$. The cases $(a)$, $(b)$ and $(c)$ correspond to $-1<n_V/n_T<-1/15$, $-1/15<n_V/n_T<0$ and $0<n_V/n_T<1$. For models of type  {\rm (I)}, an extremum exists in case $(b)$ only. For models of type {\rm (II)}, an extremum occurs in cases $(a)$ and $(b)$ i.e. $n_V<0$.}
\vspace{-.4cm}
\label{fig_V}
\end{center}
\end{figure}

Assuming an homogeneous and isotropic Universe with a flat 3-dimensional subspace as in Eq. (\ref{ansatz}), the 1-loop components of the stress-tensor are found to be:
\be
P=T^4\, p(z)\qquad \and \qquad \rho=T^4\, \big(3p(z)-p_z(z)\big):=T^4\, r(z)\, ,
\ee
where $p_z$ stands for a derivation with respect to $z$,  $p_z={\partial \over \partial z} p$.
The fields are only time-dependent and their equations of motion are:
\begin{eqnarray}
&&
\label{eq1}
3H^2={1\over 2} \dot \Phi^2+{1\over 2} \dot \phi_\bot^2+\rho\, ,\\
&&\label{eq2}
\dot \rho+3H(\rho+P)+\sqrt{3\over 2}\, \dot \Phi\, (3P-\rho)=0\, , \\
&&\label{eq3}
\ddot{\Phi}+3H\dot\Phi={\partial P\over \partial \Phi}\equiv \sqrt{3\over 2}\, (3P-\rho)\, ,\\
&&\label{eq4}
\ddot\phi_\bot+3H\dot\phi_\bot=0\qquad \Longrightarrow\qquad \dot
\phi_\bot=\sqrt{2}\; {c_\bot\over a^3}\, ,
\end{eqnarray}
where $c_\bot$ is an integration constant. Denoting $\displaystyle \o f\equiv \frac{d f}{d \ln a}$, we can first obtain the relation $\o z=\sqrt{3\over 2}\,\o \Phi-\frac{\o T}{T}$.  Differentiating this identity and using the equations of motion, we can substitute Eq. (\ref{eq3}) with an equation for $z$ of the form
\be
\label{eqz}
h(z,\o z,\o \phi_\bot)\left( \mathcal{A}(z)\oo z +\mathcal{B}(z)\o z{}^2\right)+\mathcal{C}(z)\o z+V_z(z)=0,
\ee
where we introduce the notion of an effective potential $V$:
\be
\label{hatV_z}
V_z(z)=r-4 \, p \, .
\ee
The explicit expressions for $h(z,\o z,\o \phi_\bot)$, $\mathcal{A}(z)$, $\mathcal{B}(z)$ and $\mathcal{C}(z)$ can be found in Ref.\cite{Cosmo-3}.
Eq. (\ref{eqz}) admits a static solution $z\equiv z_c$, $\phi_\bot\equiv cst.$ when the potential $V(z)$ admits a critical point $z_c$. It happens that the shape of $V(z)$ depends drastically on the parameters $n_V/n_T$ and $n'_T$:
\begin{itemize}
\item
In case (I), $n_T'=0$ and three behaviors can arise, as depicted on Fig. \ref{fig_V}:
\begin{itemize}
\item Case (I$a$): For $\dis {n_V\over n_T}<-{1\over 15}$, $V(z)$ increases.
\item  Case (I$b$): For $\dis -{1\over 15}<{n_V \over n_T}<0$, $V(z)$ has a unique minimum $z_c$, and $p(z_c)>0$.
\item  Case (I$c$): For $\dis 0<{n_V\over n_T}$, $V(z)$ decreases.
\end{itemize}
\item
In case (II), $n_T'>0$. In the three above ranges, the behaviors differ from case (I) for large negative $z$,  where the potentials are linearly decreasing (see Fig. \ref{fig_V}):
\begin{itemize}
\item Case (II$a$) \& (II$b$): For $n_V<0$, $V(z)$ has a unique minimum $z_c$, and $p(z_c)>0$.
\item  Case (II$c$): For $\dis 0<n_V$, $V(z)$ decreases.
\end{itemize}
\end{itemize}
Thus, the cases (I$b$), (II$b$) and (II$a$) admit a very particular solution where both scalars are constants \ie $z\equiv z_c$ and $c_\bot=0$. The conservation of the stress-tensor (\ref{eq2}) and the Friedmann equation (\ref{eq1}) gives then:
\be
\label{RDS2}
M(t)=T(t)\, e^{z_c}={1\over a(t)}\times a_0M_0\qquad \with \qquad a(t)=\sqrt{t\over t_0}\times a_0\; ,\quad \phi_\bot=cst.\,
\ee
where $a_0$, $M_0$, $t_0$ are constants.
This evolution is characterized by a temperature $T$, a spontaneous supersymmetry breaking scale $M$ and an inverse scale factor that are proportional for all times, with $H^2\propto 1/a^4$ and $\phi_\bot$ a modulus. It is thus an RDS.

\begin{figure}[h!]
\begin{center}
\vspace{.3cm}
\includegraphics[height=5.5cm]{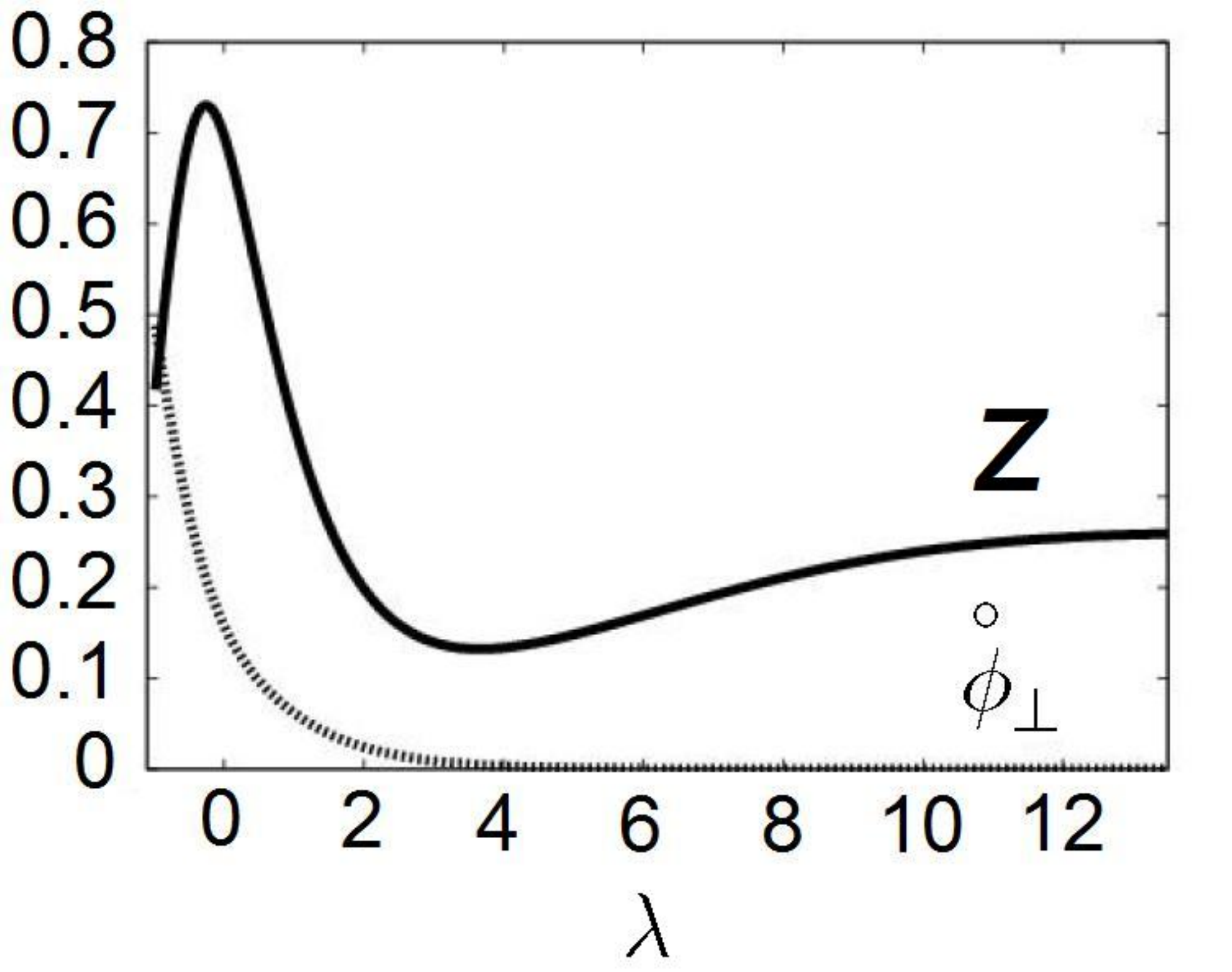}
\vspace{-.3cm}
\caption{\footnotesize \em Example of damping oscillations of $z(\lambda)$ where $\lambda\equiv \ln a$ (solid curve) and convergence to zero of $\overset{\circ}\phi_\bot(\lambda)$ (dotted curve) illustrating the dynamical attraction towards the critical solution $z\equiv z_c\simeq 0.272$, $\dot\phi_\bot\equiv 0$. It correspond to ${n_V\over n_T}=-0.02$, $n_T'=0$ i.e. some case {\rm (I$b$)}. The initial conditions are $(z_0, \overset{\circ}{z}_0, \overset{\circ}{\phi}_{\bot 0})=(0.4, 0.8, 0.5)$.}
\vspace{-.4cm}
\label{fig_oscillations}
\end{center}
\end{figure}

Using the positivity  properties  of $h(z,\o z,\o \phi_\bot)$, $\mathcal{A}(z)$, $\mathcal{B}(z)$ and $\mathcal{C}(z)$, it is easy to check analytically that this RDS is stable for small perturbations, implying that it is a local attractor of the dynamics \cite{Cosmo-3}. Global attraction is also true, but numerics are required to yield to this conclusion. Fig. \ref{fig_oscillations} gives an example of convergence to the RDS obtained for a generic choice of I.B.C..

In case (I$a$), one can show analytically that when $z\ll-1$ and $\abs \o z\abs \ll 1$, the friction due to the expansion of the Universe implies $\o z$ to converge to 0. In other words, $z$ is attracted  to an arbitrary constant along the flat region of $V(z)$. The cosmological evolution is thus converging to the RDS (\ref{RDS2}), where $z_c\ll -1$ is a modulus whose value is now determined by the I.B.C.. A numerical study shows that for arbitrary I.B.C., $z$ ends by sliding along its potential and enters the regime $z\ll -1$, $\abs \o z\abs \ll 1$ that yields to the above conclusions. It follows  that the evolution is always attracted to the RDS. However, since  $e^z=R_0/R_4\ll 1$ with $R_0\gg 1$, it is more natural to interpret the RDS from a 5-dimensional point of view \ie to consider $S^1(R_4)$ as part of the space-time itself. In that case, $R_4$ does not appear in the definition of a scalar field $M$ (or $\Phi$), but is interpreted as a fifth component of the metric, $g_{st44}=(2\pi R_4)^2$. The attractor     (\ref{RDS2}) is then rewritten as an RDS in 5 dimensions, where supersymmetry is spontaneously broken by thermal effects only: The scale factor $a'(t)$ of the directions 1, 2, 3, the scale factor $b(t)$ of the direction 4 and the temperature $T'$ of the 5-dimensional Universe evolve proportionally for all times and one has $H^{\prime 2}\propto 1/a^{\prime 5}$. The mechanism described in case (I$a$) thus corresponds to the {\em dynamical decompactification} of an internal direction involved in the spontaneous breaking of supersymmetry.

Finally, in cases (I$c$) and (II$c$), one can show that the scale factor ends by decreasing (eventually after a turning point where $\dot a=0$) and that $\rho$ and $P$ are formally diverging at late times \cite{Cosmo-3, Cosmo-4}. This implies that our underlying hypothesis of quasi-staticness of the evolution breaks down at some time, since the perturbations of the background are very large. Thermodynamics out of equilibrium should then be applied and is out of the scope of the present work.

As a conclusion, the originally static models based on internal backgrounds $\M_6$ given in Eq. (\ref{I,II}) or of the form $\dis {S^1(R_4)\times T^5\over \Z_2\times \Z_2}$ are giving rise to cosmological evolutions attracted to RDS if and only if the partition function $\Z_{\rm 1-loop}$ contains a negative contribution, $n_V\le 0$.\footnote{The limit case $n_V=0$ can be seen to yield an RDS in 4 dimensions \cite{Cosmo-3}.} In addition,  the space-time dimension of the late time evolution is determined dynamically in case (I).  It is important to mention that during the attraction to the RDS, there is no substantial period of accelerated expansion for the Universe \ie this intermediate era cannot account for inflation. Moreover, it is easy to see numerically that the time needed to reach the RDS can exceed the age of our real Universe, especially in case (I). To avoid this problem, one can start with I.B.C. close enough to the radiation era or restrict to models in case (II), due to the replacement of the plateau for $z\ll -1$ by a steeper potential. Note that the more realistic models where $\N=1$ is spontaneously broken belong precisely to this class.

\subsection{{\em n} = 2 models and complex structure stabilization}
\label{n=2mod}

New phenomena can occur when the results of the previous section are extended to models with non-trivial boundary conditions along $n=2$ internal directions, say 4 and 5. To be specific, let us analyze 4-dimensional Euclidean backgrounds with internal space
\be
\M_6= S^1(R_4) \times S^1(R_5) \times \M_4,
\ee
where $R_0, R_4, R_5 \gg 1$ to avoid any risk of Hagedorn-like phase transition. Again, we restrict ourselves to radii $R_I$ of the space $\M_4$ satisfying
\be
{1\over R_i}\ll R_I\ll R_i\qquad (i=0,4,5\, ; \quad I\neq 4,5).
\ee
As before, the dominant contribution to the 1-loop partition function $Z_{\rm 1-loop}$ arises from the pure KK excitations along the circles $S^1(R_i)$ ($i=0,4,5$), while the other modes give exponentially damped terms that can be safely neglected. By analogy with Eq. (\ref{ZTM}), $Z_{\rm 1-loop}$ can be expressed in terms of triple discrete sums involving two ratios of radii \ie two ``complex structures''. A convenient choice for them is,
\be
e^z:={R_0\over \sqrt{R_4R_5}}\; , \qquad  e^Z:={R_5\over R_4}\; .
\ee
In terms of the temperature $T$ and the supersymmetry breaking scale $M$, the free energy density is found to take the following form,
\be
\F=-T^4\, p(z,Z)\qquad \where\qquad e^z={M\over T}\; , \quad T={1\over 2\pi\,  R_0 e^{-\phi}}\; , \quad M={1\over 2\pi \sqrt{R_4R_5} \, e^{-\phi}}.
\ee

Depending on the precise way to break supersymmetry along the internal directions 4 and 5, different cosmological behaviors are found. In some cases one or the other of the directions 4 and 5 is spontaneously decompactified and we are back to a case already treated in the previous subsection (generalized in higher dimensions). In other cases, $(z,Z)$ is found to converge to a critical point $(z_c,Z_c)$ corresponding again to an RDS in 4 dimensions, where $T(t)\propto M(t)\propto 1/a(t)$. The new phenomenon encountered in such simple models is the dynamical stabilization of the complex structure $e^Z=R_5/R_4$.

%%%%%%%%%%%%%%%%%%%%%%%%%%%%%%%%%%%%%%%%%%%%
%%%%%%%%%%%%%%%%%%%%%%%%%%%%%%%%%%%%%%%%%%%%

\section{Stabilization of K{\"a}hler structures}
\label{hetpot}

In the previous sections, we have supposed that the internal radii-moduli $R_I$ that are not participating in the spontaneous breaking of supersymmetry are bounded by the radii (and their inverses) that do participate in the breaking  (see Eqs (\ref{inter}) and (\ref{inter2})). This hypothesis was fundamental to ague that the $R_I$'s appear in the 1-loop free energy through exponentially suppressed terms only. Here, we would like to justify this assumption is consistent by analyzing the dynamics of the$R_I$'s, the so-called ``\Ka moduli'', for arbitrary initial values and velocities.

Since $R_I$ can {\em a priori} be large, the associated $S^1(R_I)$ may be treated as a space-time direction rather than an internal one. We therefore consider the framework of the previous sections with arbitrary number $d-1$ of large spatial directions and, for simplicity, we consider the dynamics of a single internal circle, $S^1(R_d)$.
As an example, we introduce a spontaneous breaking of supersymmetry that involves $n=1$ internal circle, say in the direction 9, $S^1(R_9)$. Temperature is implemented as usually with non-trivial boundary conditions along the Euclidean time    $S^1(R_0)$. Altogether, we consider the following 10-dimensional background in heterotic or type II superstring,
\be
S^1(R_0)\times T^{d-1}(R_{\rm box}) \times S^1(R_d)\times \M_{10-d-2}\times S^1(R_9),
\ee
where $R_0\gg 1$ and  $R_9\gg 1$, while the remaining radii $R_I$ of the internal manifold $\M_{10-d-2}$ satisfy
\be
\label{inter3}
{1\over R_0}\ll R_I\ll R_0\qquad \and \qquad {1\over R_9}\ll R_I\ll R_9 \qquad (I\neq d).
\ee
Since we do not consider a $\Z_2$ orbifold action on $S^1(R_9)$, we are actually considering models in case (I), in the notations of section 3 (see \cite{Cosmo-4} for other cases).

Due to the T-duality on $S^1(R_d)$, $Z_{\rm 1-loop}$ admits a symmetry $R_d\to 1/R_d$ that relates the regime $R_d\ge 1$ to $R_d\le 1$. When $R_d\gg 1$, the light states are the KK modes along the directions $S^1(R_0)$, $S^1(R_9)$ and $S^1(R_d)$. As in section \ref{n=2mod}, their contribution to $Z_{\rm 1-loop}$ involves two complex structures, say $R_0/R_9$ and $R_9/R_d$. On the contrary, when $R_d$ approaches 1 from above, not only the winding but also the KK modes along $S^1(R_4)$ cease to contribute substantially, since their mass is 1 (in $\sqrt{\alpha'}$ units), up to a numerical factor. However, in heterotic string theory, this numerical factor happens to be 0 for the particular modes whose winding and momentum numbers are $\tm_d=n_d=\pm 1$ (the extended gauge symmetry point $U(1) \rightarrow SU(2)$ at $R_d=1$). In other words, while these specific states are super massive for generic values of $R_d$, their KK towers along  $S^1(R_0)$, $S^1(R_9)$ do contribute when $R_d\simeq 1$. Defining
\be
e^z:= {R_0\over R_9}\; , \qquad e^\eta:= R_9\; , \qquad e^\zeta := R_d,
\ee
and dropping terms that are exponentially suppressed in any regime of $R_d$,
the string partition is found to be
\be
Z_{\rm 1-loop}=\beta V_{\rm box}\,{1\over (2\pi R_0)^d}\,  p(z,\eta,\zeta),
\ee
where $p$ can be expressed in one way or another as,
\be
\label{p}
\begin{array}{ll}
p(z,\eta,\zeta)&\!\!\! =n_T\, \left[\hat f^{(d)}_T(z)+k^{(d)}_T(z,\eta-\abs\zeta\abs)\right]+n_V\, \left[\hat f^{(d)}_V(z)+k^{(d)}_V(z,\eta-\abs\zeta\abs)\right]\\ \\&~~+\tilde n_T\, g^{(d)}_T(z,\eta, \abs\zeta\abs)+\tilde n_V\, g^{(d)}_V(z,\eta, \abs\zeta\abs)\\ \\
&\!\!\!= e^{\abs\zeta\abs-\eta-z}\left[n_T\, f^{(d+1)}_T(z,\eta-\abs\zeta\abs)+ n_V\, f^{(d+1)}_V(z,\eta-\abs\zeta\abs)\right]\\\\
&~~+\tilde n_T\, g^{(d)}_T(z,\eta, \abs\zeta\abs)+\tilde n_V\, g^{(d)}_V(z,\eta,\abs \zeta\abs).
\end{array}
\ee
Note that $p$ is an even function of $\zeta$, as follows from T-duality $R_d \to 1/R_d$.
$n_T$ is the number of massless states for generic $R_d$, while $n_V$ is the difference between the numbers of bosons and fermions which are massless. $\tn_T$ is the number of additional massless states at the self-dual point $R_d=1$ and $\tn_V =\tn_T$ because these modes are bosons. Physically,  the KK reduction of the 10-dimensional metric tensor along $S^1(R_d)$ provides generically an $U(1)$ gauge theory, which is enhanced to $SU(2)$ when $(R_d-1/R_d)$, interpreted as a Higgs VEV, vanishes. The properties of the functions $g^{(d)}_T$ and $g^{(d)}_V$ is precisely to interpolate between the different massless spectra ($U(1)$ {\em versus} $SU(2)$). They are not functions of two complex structures $z$ and $\eta-\abs \zeta\abs$ only, since the string scale $\sqrt{\alpha'}$ is entering the game.
The definitions of the various functions appearing in Eq. (\ref{p}) are
\be
\label{fkgjT}
\begin{array}{ll}
&\hat f^{(d)}_T(z)=\displaystyle{\Gamma\left({d+1\over 2}\right)\over \pi^{d+1\over 2}}\sum_{\tk_0,\tk_9}{e^{dz}\over \left[e^{2z}(2\tk_0+1)^2+(2\tk_9)^2\right]^{d+1\over 2}}\, , \\
&k^{(d)}_T(z,\eta-\abs\zeta\abs)=\displaystyle {\sum_{m_d}}'\abs m_d\abs^{d+1\over 2}e^{{d+1\over 2}(\eta-\abs\zeta\abs)}e^{dz} \sum_{\tk_0,\tk_9}{2K_{d+1\over 2}\left(2\pi\abs m_d\abs e^{\eta-\abs\zeta\abs}\sqrt{e^{2z}(2\tk_0+1)^2+(2\tk_9)^2}\right)\over \left[e^{2z}(2\tk_0+1)^2+(2\tk_9)^2\right]^{d+1\over 4}}\, , \\
&g^{(d)}_T(z,\eta,\abs\zeta\abs)=\displaystyle \left(e^{2\abs\zeta\abs}-1\right)^{d+1\over 2} e^{{d+1\over 2}(\eta-\abs\zeta\abs)}e^{dz}\sum_{\tk_0,\tk_9}{2K_{d+1\over 2}\left(2\pi (e^{2\abs\zeta\abs}-1)e^{\eta-\abs\zeta\abs}\sqrt{e^{2z}(2\tk_0+1)^2+(2\tk_9)^2}\right)\over \left[e^{2z}(2\tk_0+1)^2+(2\tk_9)^2\right]^{d+1\over 4}}\, , \\
&f^{(d+1)}_T(z,\eta-\abs\zeta\abs)=\displaystyle{\Gamma\left({d\over 2}+1\right)\over \pi^{{d\over 2}+1}}\sum_{\tk_0,\tk_9,\tm_d}{e^{(d+1)z}\over \left[e^{2z}(2\tk_0+1)^2+(2\tk_9)^2+e^{-2(\eta-\abs\zeta\abs)}\tm_d^2\right]^{{d\over 2}+1}}\ ,
\end{array}
\ee
with the remaining ones given as
\be
\label{fkgjV}
\begin{array}{lll}
&\hat f^{(d)}_V(z)=e^{(d-1)z}\, \hat f^{(d)}_T(-z), &k^{(d)}_V(z,\eta-\abs\zeta\abs)=e^{(d-1)z}\, k^{(d)}_T(-z,\eta-\abs\zeta\abs+z), \\ \\
& g^{(d)}_V(z,\eta,\abs\zeta\abs)=e^{(d-1)z}\, g^{(d)}_T(-z,\eta+z,\abs\zeta\abs),& f^{(d+1)}_V(z,\eta-\abs\zeta\abs)=e^{dz}\, f^{(d+1)}_T(-z,\eta-\abs\zeta\abs+z).
\end{array}
\ee
In the type II case, the partition function takes formally the form of the heterotic one, with $\tilde n_T=\tilde n_V=0$. This is due to the fact that in type II, there is no enhancement of symmetry at $R_d=1$.

\begin{figure}[h!]
\begin{center}
\vspace{0.3cm}
\includegraphics[height=5.7cm]{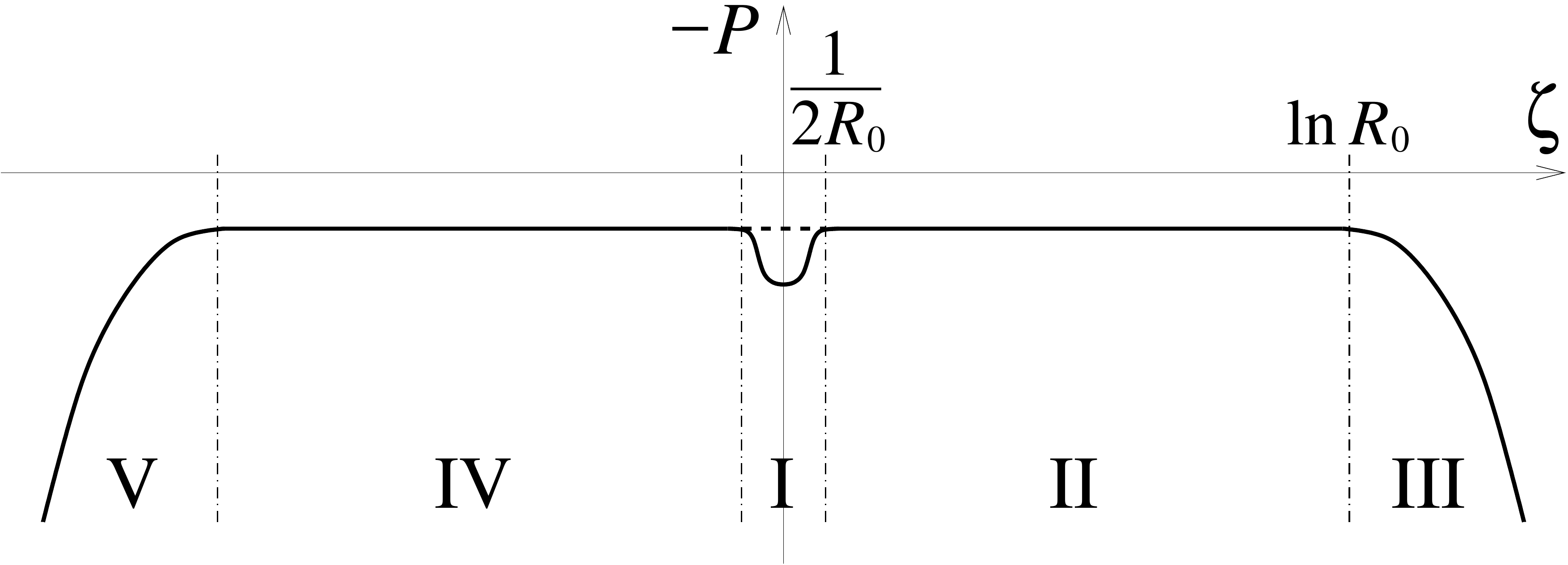}
\caption{\footnotesize \em Qualitative shape of the thermal effective potential $-P$ of $\zeta=\ln R_d$ (the other variables held fixed, with $z<0$). There are five phases in the heterotic models, while the range {\rm I} is reduced to a single point in type II. When $z>0$, one has to replace $R_0$ by $R_9$ in the boundaries of the ranges and if $z$ is large enough, phase {\rm III} ({\rm V}) is increasing (decreasing).}
\label{fig_potential}
\end{center}
\vspace{-0.5cm}
\end{figure}

Treating the circle $S^1(R_d)$ as an internal direction, the dimensional reduction from 10 to $d$ dimensions involves the metric and dilaton field $\phi_d$ in $d$ dimensions, together with the scalars $\eta$ and $\zeta$,
\be
\label{Scompact}
S=\int d^dx\sqrt{-g}\left( {R\over 2}-{1\over 2}(\partial \Phi)^2-{1\over 2}(\partial \phi_\bot)^2-{1\over 2}(\partial \zeta)^2 -\F\right),
\ee
where we have defined the normalized fields
\be
\label{Phiphibot}
\Phi:={2\over \sqrt{(d-2)(d-1)}}\, \dil-\sqrt{d-2\over d-1}\, \eta\; , \qquad \phi_\bot:={2\over \sqrt{d-1}}\, \dil +{1\over \sqrt{d-1}}\, \eta\, .
\ee
The 1-loop free energy density $\F=-Z_{\rm 1-loop}/(\beta V_{\rm box})$ depends on the temperature $T$, the supersymmetry breaking scale $M$ (\ie $\Phi$), $\zeta$ and implicitly on $\phi_\bot$ via $\eta$,
\be
\F=-T^d\, p(z,\eta,\zeta)\; , \qquad e^z={M\over T}\; ,\qquad T={e^{2\phi_d\over d-2}\over 2\pi R_0}\; , \qquad M={e^{2\phi_d\over d-2}\over 2\pi R_9}\equiv {e^{\sqrt{d-1\over d-2}\Phi}\over 2\pi}.
\ee
As usually, a FLRW ansatz for the metric and time-dependent scalars yields a stress-tensor whose thermal energy density and pressure are
\be
P=T^d\, p(z,\eta,\zeta)\qquad \and \qquad \rho=T^d\, r(z,\eta,\zeta)\; , \quad r=(d-1)p-p_z.
\ee
Among the five independent equations of motions, we are particularly interested in the equation for $\zeta$,
\be
\label{zetaEq}
\ddot \zeta+(d-1)H\dot \zeta- {\partial P\over \partial\zeta}=0,
\ee
whose potential $-P$, as a function of $\zeta$ (the over fields held fixed), is shown on Fig. \ref{fig_potential} (when $z<0$ \ie $R_0<R_9$).
In the heterotic case, the profile of $-P$ can be divided in five phases, while in type II models the range I is reduced to a single point. The behavior in phase III (phase V)  is exponentially decreasing (increasing), while when $z> 0$ and large enough, it is exponentially increasing (decreasing).

%%%%%%%%%%%%%%%%%%%%%%%%%%%%%%%%%%%%%%%%%%%%%%%
\vspace{.4cm}

\noindent {\large \em heterotic models}
\begin{itemize}

\item {\it \bf  I : Higgs phase,}  $\dis \left\abs R_d-{1\over R_d}\right\abs < {1\over R_0} ~{\rm and/or}~{1\over R_9}.$

The functions $k_{T}^{(d)}$ and $k_{V}^{(d)}$ in the first expression of Eq. (\ref{p}) are exponentially suppressed and can be neglected. Since $p$ is an even function of $\zeta$, we have at the origin $p_\zeta = 0$ so that $\zeta(t) \equiv 0$ is a solution to (\ref{zetaEq}). It follows that the pressure is drastically simplified, since
\be
\label{pzeta0}
p(z,\eta,0)= (n_T+\tilde n_T)\, \hat f_T^{(d)}(z)+(n_V+\tilde n_V)\, \hat f_V^{(d)}(z):= \tilde p(z)\, ,
\ee
\ie
does not depend on $\eta$. Thus, the analysis of the particular solution $\zeta\equiv 0$ brings us back to the study of section \ref{n=1mo}, once generalized in $d$ dimensions. We conclude that when
\be
- {1 \over 2^d - 1} < {n_V + \tilde n_V \over n_T + \tilde n_T} < 0,
\ee
there exists an RDS$^d$ (Radiation Dominated Solution in $d$ dimensions). The behavior of the first order fluctuations around this solution are found to be exponentially damped. As a result, the RDS$^d$ is a local attractor of the dynamics and $R_d$ is stabilized at the self-dual point of enhanced symmetry.

\item {\it \bf II : Flat potential phase,} $\dis {1\over R_0}~ {\rm and} ~{1\over R_9}< R_d - {1\over R_d}  < R_0~ {\rm and} ~R_9.$

As in phase I, $k_{T}^{(d)}$ and $k_{V}^{(d)}$ in Eq. (\ref{p}) can be neglected. Indeed, by definition, phase II starts when the functions $g_{T}^{(d)}$ and $g_{V}^{(d)}$ responsible for the interpolation between the generic and the enhanced massless spectra are exponentially suppressed as well. It follows that $p$ is independent of $\eta$ and $\zeta$,
\be
\label{pplat}
p(z,\eta,\zeta)\simeq n_T\, \hat f_T^{(d)}(z)+n_V\, \hat f_V^{(d)}(z):=\hat p(z)\, .
\ee
Consequently, any $\zeta(t)\equiv \zeta_0$ when $\zeta$ is in the range II solves Eq. (\ref{zetaEq}). For any given $\zeta_0$, we are back again to the analysis of section \ref{n=1mo} and a particular RDS$^d$ exists when
\be
- {1 \over 2^d - 1} < {n_V \over n_T} < 0.
\ee
Small perturbations around such an RDS$^d$ are found to the damped, even if the potential for $\zeta$ is flat. Actually, the fluctuations of $\zeta$ around any $\zeta_0$ are suppressed due to  the presence of ``gravitational friction" in (\ref{zetaEq}). In this sense, one can conclude that $\zeta$ is marginally stabilized.

\item  {\it \bf III : Higher dimensional phase,} $R_0~ {\rm and/or} ~R_9  <R_d.$

As in phase II, $g_{T}^{(d)}$ and $g_{V}^{(d)}$ in Eq. (\ref{p}) can be neglected. However, by definition, phase III starts when the functions $k_{T}^{(d)}$ and/or $k_{V}^{(d)}$ cease to be exponentially suppressed. In particular, when $R_d\gg R_0$ and $R_9$, one finds, using the second form of Eq. (\ref{p}),
\be
\label{pzetainf}
p(z,\eta,\zeta)\simeq e^{\abs\zeta\abs-\eta-z}\left(n_T\, \hat f_T^{(d+1)}(z)+n_V\, \hat f_V^{(d+1)}(z)+e^{d(z+\eta-\abs \zeta\abs)}(n_T+n_V){S^{oe}_d\over 4}\right)\, ,
\ee
where $S_d^{oe}={\Gamma\left({d\over 2}\right)\over \pi^{d\over 2}}\, \underset{m} {\sum}'{1\over \abs m\abs^d}$. The term $\dis e^{d(z+\eta-\abs \zeta\abs)}= (R_0/ R_d)^d$  is power-like subdominant and exponentially small terms have been ignored.

Neglecting the small contribution $(R_0/ R_d)^d$, the appearance of the functions $\hat f_T^{(d+1)}$ and $\hat f_V^{(d+1)}$ in $p$ indicate that it is more natural to reconsider the system from a $(d+1)$-dimensional point of view. Regarding $S^1(R_4)$ as part of the space-time, $R_4$ is no longer an internal modulus but a component of the metric, $g_{st,dd}=(2\pi R_d)^2$. Denoting with primes all quantities in $d+1$ dimensions, the vacuum-to-vacuum energy density in the effective action in higher dimension, $Z_{\rm 1-loop}/(\beta V_{\rm box} 2\pi R_d)$, is giving rise to can a pressure  $P'=T^{\prime (d+1)}\, p'(z,\eta,\zeta)$, where
\be
p'(z,\eta,\zeta) \simeq n_T\, \hat f_T^{(d+1)}(z)+n_V\, \hat f_V^{(d+1)}(z):=\hat p'(z)\, .
\ee
One more time, one concludes from the analysis of section \ref{n=1mo} that an RDS$^{d+1}$ with isotropic metric exists if
\be
- {1 \over 2^{d+1} - 1} < {n_V \over n_T} < 0.
\ee
By isotropic metric, we mean that
\be
e^\xi:= b/a'=R_d/R_{\rm box},
\ee
where $a'$ is the scale factor in the directions $1,\dots, d-1$ and $b$ is the scale factor in the direction $d$, satisfies $\xi(t)\equiv \xi_0$, a constant determined by the I.B.C.. Small perturbations around such a solution are shown to converge to zero. Since $\xi(t)$ measures the anisotropy of the local metric, one concludes that the attracting RDS$^{d+1}$ is characterized by an enhanced spatial rotation group $SO(d-1)\to SO(d)$. The ratio $e^{\xi_0}$ is interpreted as a ``complex structure'' of the ``external space'' we live in.

On the contrary, whenever the contribution $(R_0/R_d)^d$ in Eq. (\ref{pzetainf}) is not negligible, we find it yields a ``residual force'' such that even though $\zeta(t)$ \ie $R_d(t)$ is increasing, it is always caught by $R_9(t)$ and $R_0(t)$. Thus, the dynamics exits phase III and the system enters into region II, where the solution is attracted to an RDS$^d$.

\item {\it \bf IV : Dual flat potential phase,} $\dis {1\over R_0}~ {\rm and} ~{1\over R_9}<  {1\over R_d} -R_d  < R_0~ {\rm and} ~R_9.$

This region is the T-dual of phase II and has the same behavior, with $\zeta \rightarrow - \zeta$. The light states that contribute to $\Z_{\rm 1-loop}$ are the windings modes along $S^1(R_d)$ instead of the KK excitations.

\item {\it \bf V :  Dual higher dimensional phase,} $\dis  R_0~ {\rm and/or} ~R_9<{1\over R_d}. $

This region is the T-dual of phase III and has the same behavior, with $\zeta \rightarrow - \zeta$.

\end{itemize}

To summarize, when the \Ka modulus $R_d(t)$ is internal, it is attracted to the intersection of the ranges $1/R_i(t)<R_d(t)<R_i(t)$ $(i=0,9)$ and ends by being marginally stabilized or stabilized at $R_d=1$. On the contrary, when $R_d(t)$ is large enough, the Universe is $(d+1)$-dimensional and $R_d$ expands and runs away, with $R_d(t)\propto R_{\rm box}(t)$. It is then better understood in terms of the complex structure $R_d/R_{\rm box}$, which is marginally stabilized.

%%%%%%%%%%%%%%%%%%%%%%%%%%%%%%%%%%%%%%%%%%%%%%%
\vspace{.4cm}

\noindent {\large \em type II superstring models}

\noindent As said before, there is no enhanced symmetry point at $R_d=1$ in the type II superstring models. Thus, their analysis can be derived from the heterotic one by taking $\tilde n_T=\tilde n_V=0$. Since the local minimum of $-P$ at $\zeta=0$ is not present anymore, the plateaux II and IV on Fig. \ref{fig_potential} are connected.

However, we expect by heterotic-type II duality that an Higgs phase I should exist in type II superstring at the non-perturbative level. A possible setup to describe this effect is to consider a pair of D-branes, whose distance is dual to the modulus $R_d$. In this context, our Universe is a ``brane-world'', whose spatial directions are parallel to the D-branes. The stabilization of $R_d$ at the self-dual point on the heterotic side should imply  the non-perturbative thermal effective potential in type II to force the D-branes to stay on top of each other, thus producing an $U(1) \rightarrow SU(2)$ enhancement. However, this attraction between the D-branes should only be local, since if they are separated enough so that the dual modulus $R_d$ enters phase II, the thermal effective potential should allow stable finite distances between the D-branes. If the distance between the D-branes is very large, the force between them will be repulsive and the expanding Universe develops one more dimension.

An alternative non-perturbative type II set up realizing a gauge group enhancement involves singularities in the internal space. For example, a type IIA D2-brane wrapped on a vanishing 2-sphere whose radius is dual to $R_d$ produces an $SU(2)$ gauge theory. It also admits a mirror description in type IIB \cite{het/II}. The equivalence between the brane-world and geometrical singularity pictures can be analyzed along the lines of Ref. \cite{KP}.

%%%%%%%%%%%%%%%%%%%%%%%%%%%%%%%%%%%%%%%%%%%%%%%%
%%%%%%%%%%%%%%%%%%%%%%%%%%%%%%%%%%%%%%%%%%%%%%%%

\section{Conclusions and Perspectives}
%It is time to summarize our results and elaborate on further work.

In this review, we describe some basics of early Universe cosmology in the framework of string theory. We first place the $(d-1)$-dimensional space in a very large box, while much smaller compactified directions span an internal space. To introduce temperature, we consider the Euclidean version of the background, where appropriate boundary conditions are imposed along the Euclidean time circle. Supersymmetry breaking is implemented in a similar way by choosing non-trivial boundary conditions along internal compact directions. In four dimensions, we analyze models where $\N=4,2,1$ supersymmetry is spontaneously broken. 
The advantage of string theory is to cure all UV diverges encountered in field theory.  

In this work, we restrict our study to the intermediate times $t_E\ll t\ll t_{W}$ \ie after the end of the Hagedorn era  and before the electroweak phase transition. 
The pressure and energy density of the gas of strings are computed from a microscopic point of view, using the 1-loop Euclidean string partition function. We then study the solutions of the low energy effective action and find that the quasi-static evolutions are attracted to Radiation Dominated Solutions. During the convergence to an RDS, there is no inflation (or a very tiny amount). However, interesting effects can occur. Internal radii that are participating into the spontanoues breaking of supersymmetry can be decompactified dynamically and lead to a change in the dimension of space-time. Moreover, the internal radii that are not participating in the spontaneous breaking of supersymmetry are stabilized. In general, the Universe tends to increase spontaneously its symmetries: The gauge symmetries or the local isotropy. 
Usually, the field theory description of the full evolution is in term of a  succession of {\em different} field theories. The underlying string theory is required to connect them.

To go further, lots of work is still needed to unravel some experimental predictions from our framework. In four dimensions, models with spontaneously broken $\N=1$ supersymmetry are particularly interesting, since they can include chiral matter. Dealing with them is under progress, in order to go beyond $t_{W}$. To do that, one needs to compute the radiative corrections to the different fields entering the action. Only after this full work is accomplished, it becomes possible to discuss what happens in the matter dominated era, and observe if a late time inflation era can exist. There are strong beliefs that $\N=1$ models will produce a non zero cosmological constant.

Another direction of work concerns the relaxation of the homogeneity assumption. We could thus deal with the issues of entropy production and adiabaticity. On simpler grounds, it could be interesting  to show that small initial homogeneities disappear and that the standard FLRW ansatz is an attractor in the space of backgrounds. 

Finally, an important issue to examine is the existence of {\it a mechanism alternative to inflation} during the Hagedorn era. There are already some proposals in this direction in Ref. \cite{BDB}. Preliminary results by some of the authors indicate  that such a  mechanism is plausible for non-pathological sting vacua, where the Hagedorn transition is resolved \cite{Carlo, MSDS}.

\section*{Acknowledgements}

The authors would like to thank NOVA publishers for giving us the opportunity to write this review.
We are grateful to N. Toumbas for useful discussions.  H.P. thanks the Ecole Normale Sup\'erieure for hospitality.\\
\noindent This work is partially supported by the ANR (CNRS-USAR) contract 05-BLAN-0079. The work of F.B, J.E. and H.P is supported by the European contracts PITN-GA-2009-237920, ERC-2008-AdG 20080228, the CNRS PICS contracts  3747, 4172 and the INTAS grant 03-51-6346. J.E. acknowledges financial support from the Groupement d'Int\'er\^et Scientifique P2I.

\end{document}